\documentclass[11pt,a4paper]{article}
\usepackage{jheppub}

\usepackage{graphicx}
\usepackage{color}
\usepackage{mathtools} 
\usepackage{caption}
\usepackage{subcaption}
\usepackage{float}
\usepackage{slashed}
\usepackage{amsmath,amsfonts,graphicx,multicol}

\DeclareGraphicsExtensions{.pdf,.png,.jpg}
\newcommand\rep\mathbf
\newcommand{\newc}{\newcommand} 

\newcommand{\nn}{\nonumber}
\newc{\barr}{\begin{eqnarray}}
\newc{\earr}{\end{eqnarray}}
\newc{\dis}{\displaystyle}
\newc{\beq}{\begin{equation}}
\newc{\eeq}{\end{equation}}

\begin{document}

 \title{NLO QCD corrections to the resonant Vector Diquark production at the LHC}
\author[a]{Kasinath Das}
\author[b]{Swapan  Majhi}
\author[a]{Santosh Kumar Rai}
\author[a]{Ambresh Shivaji}

\affiliation[a]{Harish-Chandra Research Institute and  
                Regional Centre for Accelerator-based Particle Physics, \\
                Chhatnag Road, Jhusi, Allahabad 211019, India}
\affiliation[b]{Department of Theoretical Physics,  
                Indian Association for the Cultivation of Science \\
                2 A \& B Raja S C Mullick Road,
                Kolkata 700032, India} 

\emailAdd{kasinathdas@hri.res.in}
\emailAdd{majhi.majhi@gmail.com}
\emailAdd{skrai@hri.res.in}
 \emailAdd{ambreshkshivaji@hri.res.in}   
%
%

\preprint{RECAPP-HRI-2015-010}
 
\abstract{With the upcoming run of the Large Hadron Collider (LHC) at 
much higher center of mass energies, the search for Beyond Standard 
Model (BSM) physics will again take center stage. New colored particles 
predicted in many BSM scenarios are expected to be produced with large 
cross sections thus making them interesting prospects as a doorway to 
hints of new physics. We consider the resonant production of such a 
colored particle, the diquark, a particle having the quantum number of 
two quarks. The diquark can be either a scalar or vector. We focus on 
the vector diquark which has much larger production cross section compared 
to the scalar ones. In this work we calculate the next-to-leading order (NLO) 
QCD corrections to the on-shell vector diquark production at the LHC 
produced through the fusion of two quarks as well as the NLO 
corrections to its decay width. We present full analytic results for 
the one-loop NLO calculation and do a numerical study to show 
that the NLO corrections can reduce the scale uncertainties in the 
cross sections which can be appreciable and therefore modify the 
expected search limits for such particles. We also use the dijet result 
from LHC to obtain current limits on the mass and coupling strengths 
of the vector diquarks.}
 

 
\maketitle

\section{Introduction}
After the successful running of the Large Hadron Collider (LHC) at CERN 
with 7 and 8 TeV center of mass energies, the data released by the two 
experiments, ATLAS and CMS have not only improved on the limits set by 
the Tevatron experiments on any new physics scenario,  but has also 
started giving some insights into the TeV scale. In addition to the observation 
of a scalar resonance at 125 GeV  \cite{Aad:2012tfa,Chatrchyan:2012ufa}   
consistent with that of the Standard Model (SM) like Higgs boson, the results 
are also in very good agreement
with predictions from the SM, with not much 
deviation. This means that the LHC data is already pushing the energy frontier 
of any Beyond Standard Model (BSM) physics predictions. However with the 
upgraded run of LHC at center of mass energy of 13 TeV and subsequently 
14 TeV, the search for new physics is expected to be more robust and as 
envisaged for the LHC run. As expected and observed from the previous 
LHC runs, the data would be most sensitive to the strongly interacting 
sector through production of new colored states. Since the initial states 
at hadron colliders such as the LHC are colored particles, the most 
dominant contributions would be through new colored resonances.  Such 
colored particles are predicted in many class of BSM theories. Resonant 
s-channel production at LHC can happen for  {\it squarks} in R-Parity 
violating supersymmetric theories \cite{Barbier:2004ez},  {\it 
diquarks}  in super-string inspired $E_6$ grand unification models 
\cite{Hewett:1988xc} or models with extended gauge 
symmetries \cite{Chakdar:2013haa,Chakdar:2012vv,Li:2012zy}, 
color-octet vectors such as {\it axigluons} 
\cite{Frampton:1987dn, Bagger:1987fz} and {\it colorons} 
\cite{Hill:1991at, Hill:1993hs, ddn:1995pr,Sayre:2011ed}, models with 
color-triplet \cite{Babu:2006wz}, color-sextet \cite{Pati:1974yy, 
Mohapatra:1980qe, Chacko:1998td} or color-octet scalars 
\cite{Hill:2002ap, Dobrescu:2007yp}. The absence of any such 
observation in the existing data put strong limits on such particle masses, 
from pair production of such states, or more strongly from resonant 
searches of new physics exchanged in the s-channel \cite{Harris:2011bh}.

These resonant colored states are most likely to decay to two light 
jets leading to not only the modification of the dijet differential 
cross section at large invariant mass but also show up as a bump in its invariant mass
distribution. Such a signal will not go unnoticed and will be fairly very 
distinct at large invariant mass values, as the significantly huge QCD 
background falls rapidly for large dijet invariant mass. Both ATLAS 
and CMS Collaborations have looked at the dijet signal and already put 
strong constraints on the mass of such 
resonances \cite{Aad:2010bc, Aad:2011bc, Aad:2014bc,Khachatryan:2010jd, Chatrchyan:2011ns, Khachatryan:2015jd}.  
We should however note that the production of such colored particles will be beset with significant 
contributions from QCD corrections, and therefore it becomes important 
to understand how much the leading order (LO) rates might change once 
these corrections are included. One finds that there have 
been significant efforts in this direction to study the next-to-leading 
order (NLO) QCD effects on production of some of the new colored particles 
\cite{Han:2009ya,Chivukula:2011ng,Chivukula:2013xla} arising in BSM at the LHC. 
Here we are interested in particular with particles of the ``diquark" type 
which carry non-zero baryon number and couple to a pair of quarks or 
anti-quarks.  The fact that LHC being a proton-proton collider will have 
valence quarks in much abundance compared to the anti-quarks, helps in 
producing the diquark as a resonance through $qq$ fusion. A lot of studies 
carried out at LO exist in the literature for such 
diquarks and  their resonant effects in the dijet 
signal \cite{Atag:1998xq, Arik:2001bc, Cakir:2005iw,Mohapatra:2007af, 
Berger:2010fy,Han:2010rf, Giudice:2011ak, Richardson:2011df}, pair 
production of top quarks \cite{Barger:2006hm, Frederix:2007gi, Zhang:2010kr, 
Kosnik:2011jr} and single top quark production at the 
LHC \cite{Gogoladze:2010xd,Karabacak:2012rn}. The one-loop NLO correction 
for scalar diquark production was considered in Ref~\cite{Han:2009ya}. We focus on 
the case of vector {\it diquarks} which are either antitriplets or sextets 
of $SU(3)_C$. Such particles will also be copiously produced as 
s-channel resonances  with much larger cross section compared to the 
scalar ones. Once produced, the vector diquark will decay and would thus contribute 
to the dijet final state or to final states involving the third-generation quarks. 

For our study of estimating the NLO corrections to the on-shell production of a 
vector diquark at the LHC, we follow in part the methodology used in 
Ref~\cite{Han:2009ya} to present our results. In Sec. \ref{sec:formalism} 
we present the formalism and give the basic interaction Lagrangian relevant 
for our study and in Sec. \ref{sec:pheno} we discuss the on-shell production 
cross section of the vector diquark, and present our calculations 
and analytic expressions for the NLO QCD results. In Sec. \ref{sec:decay} we 
give results for the one-loop corrections to the decay width of the vector diquark.  
In Sec. \ref{sec:numerics}  we give our numerical results for the NLO cross sections and 
its dependence on the choice of scale for the production of the vector diquark in 
different channels at the LHC. We also consider its effect on the experimental 
limits for such particles and finally in Sec. \ref{sec:concl} we give our conclusions 
with future outlook. Some relevant formulas are collected as an Appendix.

\section{Formalism} \label{sec:formalism}
We are interested in new colored particles that couple to a pair of 
quarks directly and carry exotic baryon number. With the LHC being a 
proton-proton machine, the initial states comprised of the the valence 
quarks ($u,d$) would lead to enhanced flux in the parton distributions 
for the collision between a pair of valence quarks such as $uu, dd$ or 
$ud$. Any new particle that couples to these pairs would carry a baryon 
number $B=\dfrac{2}{3}$ and will be charged under the SM color gauge 
group $SU(3)_C$. Such states are generally referred to as {\it 
diquarks}. These colored diquarks can be either color antitriplets or 
sextets of $SU(3)_C$. We can describe the vector diquarks following 
Ref~\cite{Karabacak:2012rn} according to color representation ($\bar{\rep3},\ \rep6$) 
and electric charge ($4/3,\ 2/3,\ 1/3$)  as 
$V_{2\cal U}^{N_D},V_{\cal U}^{N_D},V_{\cal D}^{N_D}$, where the subscripts $2{\cal U}, {\cal U}$, and 
${\cal D}$ in the fields indicate their electric charge $|Q|$ of two up type 
quarks, one up and one down type quark respectively, while $N_D(=3 ~ (6))$ is the 
dimension of  the  antitriplet (sextet) representation. 
The relevant interactions of the quarks with the different vector 
diquarks is given by the Lagrangian
\begin{equation}
\begin{multlined}
\mathcal{L}_{qqD}^{V} =  K_{ab}^{j} \Big[  \frac{\lambda_{\alpha \beta}^{2{\cal U}}}{\sqrt {1+\delta_{\alpha\beta}}} V^{j \mu}_{2{\cal U}}     
                                      \overline{{\cal U}^c}_{\alpha a} \gamma_{\mu}P_\tau {\cal U}_{\beta b} 
  + \frac{\lambda_{\alpha \beta}^{\cal U}}{\sqrt {1+\delta_{\alpha\beta}}} V_{\cal U}^{j\mu} \overline{ {\cal D}^c}_{\alpha a}\gamma_{\mu} P_\tau  {\cal D}_{\beta b}\\
  +  \lambda_{\alpha\beta}^{\cal D} V_{\cal D}^{j\mu} \overline{ {\cal U}^c}_{\alpha a} 
\gamma_{\mu}P_{\tau} {\cal D}_{\beta b}\Big]+\mathrm{h.c.} 
\label{eq:Lagrn-QQD}
\end{multlined}
\end{equation}
where $P_{\tau}=\frac{1}{2}(1\pm\gamma_{5})$ with $\tau=L,R$ 
representing left and right chirality projection operators and 
superscript $ \mu $ is the Lorentz four vector index. The $K^j_{ab}$ 
are $SU(3)_{C}$ Clebsch-Gordan coefficients with  the quark color 
indices $a,b=1-3$, and the diquark color index $j=1-N_D$,    $C$ 
denotes charge conjugation, while $\alpha,\beta$ are the fermion 
generation indices. The color factor $K^j_{ab}$ is symmetric 
(antisymmetric) under $ab$ for the ${\rep{6}}\ (\bar{\rep{3}})$ 
representation. A more general form of the Lagrangian can be found 
in Ref~\cite{Han:2010rf}. A factor of $1/\sqrt{2}$ in 
the interaction terms involving same quark flavors is introduced to keep the  
expressions for the production cross section as well as the decay width same for both different flavor and same flavor cases.
To calculate the QCD corrections to the diquark 
production, we also need to know how the vector diquark ($V_i^{\mu}$) 
interacts with the gluons, which is given by the Lagrangian\footnote{There may exist anomalous terms in the Lagrangian allowed by gauge invariance, similar to that for vector leptoquarks \cite{Blumlein:1996qp}. For simplicity, we have neglected such anomalous contributions in the gluon-diquark-diquark interaction.}, 
\begin{eqnarray}
  {\cal L}^V_{GDD} &=&  -\frac{1}{2} (V_{i\mu\nu})^\dagger (V_i^{\mu\nu}) 
                -i g_s \; V_{i\mu}^\dagger T^A_{ij}V_{j\nu}G^{A,\mu\nu}
                \label{eq:Lagrn-GDD}
\end{eqnarray}
 where, 
 \begin{eqnarray}
  V_i^{\mu\nu} &=& D^\mu_{ij} V_j^\nu - D^\nu_{ij} V_j^\mu \\
  G^A_{\mu\nu} &=& \partial_\mu G^A_\nu - \partial_\nu G^A_\mu + g_s f^{ABC} G^B_\mu G^C_\nu \\
 D_{\mu,ij} &\equiv& \delta_{ij} \partial_\mu  -i g_s \; G^A\mu T^A_{ij}.
 \end{eqnarray}
The indices  $i$ and $j$ again run from $1 \to N_D$, where $N_D$ is the 
dimension of the diquark representation. The index $A$ runs from $1 \to 
8$ and $T^A_{ij}$ are the $SU(3)_C$ generators in the diquark representation. 
Note that we have suppressed the electric charge index ($2\cal U,U,D$) 
for the diquark as we are interested only in the QCD corrections.  The 
Feynman rules for three-point vertices involving vector diquark are 
given in Appendix~\ref{app:FR}.

The diquark can couple to the initial state valence partons coming 
from both the protons, and the production of the diquark would get 
significant enhancement due to the large flux of the valence quarks in 
the proton. Therefore the production rates are only constrained by 
its coupling strength to the pair of initial quarks and its mass, 
which are the two free parameters in our analysis. Moreover, it is also 
equally probable that the vector diquarks have generation dependent 
couplings following Eq. \ref{eq:Lagrn-QQD}. Therefore the couplings 
($\lambda^{2\cal U},\lambda^{\cal U},\lambda^{\cal D}$) involved in Eq. \ref{eq:Lagrn-QQD} 
are completely arbitrary and can in principle be large. Note that most 
of them are tightly constrained by flavor physics as they might mediate 
light meson or hadron decays \cite{Barbier:2004ez, Han:2009ya}. 
Therefore the constraints on the interaction of the vector diquark 
with the lighter quarks (first and second generation) are much more 
stronger, which means that vector diquark production at the LHC can 
have different allowed interaction strengths depending on the initial 
quarks participating in the production. To make our analysis more general 
we therefore choose to present our results normalized to the coupling 
strength. Where applicable, we would also assume that we work in the 
minimal flavor violating (MFV) scheme \cite{D'Ambrosio:2002ex} for the couplings involving both the 
left- and right-chiral quarks with the vector diquark. It is worth noting that these 
colored states do not have direct coupling to a pair of gluons and thus the 
production cross section for diquark is limited by the flux of the initial partons 
in the proton at the LHC. However large QCD corrections can significantly alter 
the rates and modify the existing constraints on the mass and interaction 
strengths of such colored states. In this work we have chosen to ignore any 
electroweak corrections as interactions of the vector diquark to the electroweak 
gauge bosons might be model dependent.

\section{Production cross section at next-to-leading order}\label{sec:pheno}
\begin{figure}[t]
 \begin{center}
 \includegraphics[width=0.4\linewidth,height=0.25\linewidth]{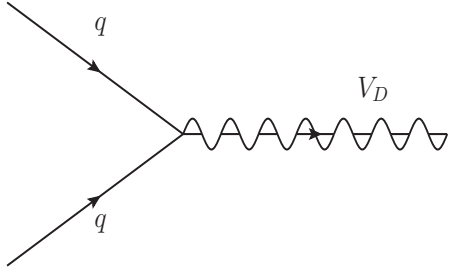}
 \end{center}
 \caption{Feynman diagram at Born level for the process $qq\rightarrow V_D.$}
  \label{fig:lo-feyn}
 \end{figure}

We shall work in the ``narrow-width" approximation where we can write the cross section as a product of 
the on-shell production and decay of the vector diquark ($V_D$) in a particular channel $(XX)$ as  
\begin{equation}
\sigma \left( p p \to XX \right) \simeq \sigma \left( p p \to V_D \right) \times \frac{\Gamma(V_D \to XX)}{\Gamma (V_D \to all)}
\end{equation}
Thus $ \sigma \left( p p \to V_D \right)$  gives the cross section for the production of the diquark resonance.
The leading order or Born contribution to the on-shell vector diquark 
production comes from quark-quark initial states. The relevant Feynman 
diagram is shown in Fig.\ref{fig:lo-feyn}. 
For the diquark of mass $M_D$, the parton-level cross section at the LO is given by
 \begin{eqnarray}
  \hat{\sigma}_{\rm B} &=& \frac{\hat{\sigma}_0}{\hat{s}}\; \delta(1-\tau), 
\label{Born}
 \end{eqnarray}
  where 
  \begin{equation}
  \hat\sigma_0 = \frac{\lambda^2 \pi N_D}{2N_C^2}. 
 \label{eqn:sigma_0}
  \end{equation}
In the above, 
$\hat{s}$ is the partonic center of mass energy, $N_C=3$ is the color factor of the quarks and 
$\tau=M_D^2/\hat{s}$. It is useful to rewrite the LO cross section in $n=4-2\epsilon$  dimensional 
form as this $n$-dimensional result will be used in the NLO calculation. Thus Eq. \ref{eqn:sigma_0} 
can be put in the form 
  \begin{equation}
  \hat\sigma_0 = \frac{(n-2)\pi N_D \lambda^2(\mu^2)\mu^{2\epsilon}}{4N_C^2}. 
 \label{eqn:sigma_LO}
  \end{equation}
Here  $\lambda(\mu^2)$ represents the running coupling parameter and $\mu$ defines the  
scale introduced to make the coupling dimensionless. From here onwards we shall drop 
the various indices  from the coupling parameter introduced in the Lagrangian~\ref{eq:Lagrn-QQD}.
The corresponding hadronic cross section 
at colliders can be obtained by convoluting the parton-level cross section with the parton 
distribution functions (PDF) of the initial quarks participating in the production, {\it i.e.}
\begin{equation}
 \sigma_{\rm LO} =  \frac{\hat \sigma_0}{s}~(q \otimes q)(\tau_0), 
 \end{equation}
 where $s$ is the hadronic center of mass energy and $\tau_0 = M_D^2/s$. We have used 
 the notation for convolution of two functions, defined by
\begin{equation}
 (f_1\otimes f_2)(x_0) = \int_0^1 dx_1\int_0^1 dx_2~ \delta(x_1x_2-x_0) f_1(x_1) f_2(x_2).
\end{equation}
%

Although the LO process involves colored particles only, the interaction strength does not involve the
strong coupling $g_s$ but only the coupling strengths given by the free parameter $\lambda$.  
Therefore the one-loop QCD corrections at NLO are in leading order of $\alpha_s=\frac{g_s^2}{4\pi}$. 
The ${\cal O}(\alpha_s)$ QCD correction to the vector diquark production involves : 
\begin{itemize}
\item Virtual corrections due to one-loop gluon contributions.  
\item Real corrections due to the gluon emission from initial state quarks and final state diquark. 
\item For the complete ${\cal O}(\alpha_s)$ correction, one also needs to consider quark-gluon initiated diquark production with a jet. 
\end{itemize}
We use dimensional regularization (DR) to regulate the ultraviolet (UV) and infrared (IR) singularities that may appear in these corrections. 
The renormalization of UV singularity and factorization of collinear singularity is carried out in the $\overline{\rm MS}$ scheme. 
We have performed various checks, including the gauge invariance check with respect to the gluon at the amplitude and 
amplitude-squared levels, to ensure the correctness of our calculations.
\begin{figure}[ht]
 \begin{center}
 \includegraphics[width=1.0\linewidth,height=0.2\linewidth]{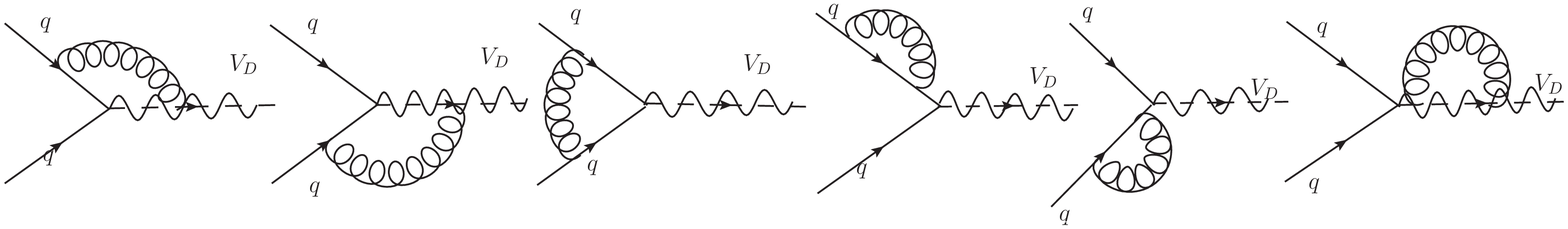}
 \end{center}
 \caption{Feynman diagrams for virtual gluon correction to the process  $qq\rightarrow V_D.$}
 \label{fig:nlo-virtual-feyn}
 \end{figure}

\subsection{Virtual corrections}
The virtual corrections at ${\cal O}(\alpha_s)$ come from the interference of Born and one-loop 
amplitudes. The one-loop diagrams contributing to virtual corrections 
are displayed in Fig.~\ref{fig:nlo-virtual-feyn}. These diagrams are 
both UV and IR divergent. The required one-loop computation is carried out following
the standard method of one-loop tensor reduction in $n=4-2\epsilon$ dimensions. We have 
listed all the one-loop scalar functions that we have used in the calculation, in Appendix~\ref{app:OLS}.
The virtual cross section coming from vertex 
correction diagrams is given by\footnote{We can also use this result to 
extract the vertex renormalization constant,
 \begin{eqnarray}
Z_{\lambda} &=& 1 - \frac{\alpha_s}{4\pi}C_\epsilon\; \Big({8\over 3}C_D+C_F\Big) \frac{1}{\epsilon_{\rm UV}}. \nn
\end{eqnarray}
}, 
\begin{equation}
\begin{multlined}
  \hat{\sigma}_{\rm V} = \hat{\sigma}_{\rm B} \frac{\alpha_s C_{\epsilon}}{2\pi} 
\Big[
C_D\Big\{ \frac{8}{3}{1\over \epsilon_{\rm UV}}  - \frac{2\pi^2}{3}+ {77\over 18}\Big\} 
+ C_F\Big\{\frac{1}{\epsilon_{\rm UV}} -\frac{2}{\epsilon_{\rm IR}^2} -\frac{4}{\epsilon_{\rm IR}} +\pi^2-8 \Big\}\Big]. 
\end{multlined}
\end{equation}
The overall factor $C_\epsilon = \frac{1}{\Gamma(1-\epsilon)} \Big( 
\frac{4\pi \mu^2}{\hat{s}}\Big)^\epsilon$ appears in all one-loop 
integrals regulated in DR. $C_F$ and $C_D$  are the eigenvalues of the 
quadratic Casimir operator of $SU(3)_C$ acting on the fundamental representation and on the 
diquark representation respectively. For both the sextet and antitriplet diquark, $C_F=4/3$ 
while  $C_D$  is $4/3$ for the antitriplet and $10/3$ for the sextet diquark. 
The effect of external leg corrections can be incorporated in the wave function renormalization of 
the quark and diquark fields. Thus one can conveniently express the sum of 
Born and virtual cross section to ${\cal O}({\alpha_s})$ as \cite{muta}, 
\begin{equation}
  \hat{\sigma}_{\rm {B+V}} =(Z_2^q)^2 Z_2^{D}\hat{\sigma} _{\rm B}+\hat{\sigma} _{\rm V}.
\end{equation}
The wave function renormalization constants $Z_2^q$ and $Z_2^D$ for 
quark and vector diquark fields are,
\begin{eqnarray}
  Z_2^q &=& 1 + \frac{\alpha_s}{4\pi}C_\epsilon\;C_F \Big( -\frac{1}{\epsilon_{\rm UV}} + \frac{1}{\epsilon_{\rm IR}}\Big) \\
  Z_2^{D} &=& 1 + \frac{\alpha_s}{4\pi}C_\epsilon\;C_D \Big( \frac{2}{\epsilon_{\rm UV}} - \frac{2}{\epsilon_{\rm IR}}\Big).
\end{eqnarray}
Note that these renormalization constants are calculated for on-shell 
quark and diquark fields, therefore, the IR singularity also appears. In 
DR, both are one as $\epsilon_{\rm UV} = \epsilon_{\rm IR}= \epsilon$. 
However, the above form is suitable for extracting the full UV singularity 
in virtual corrections. The sum of Born and virtual cross section thus becomes, 
\begin{equation}
\begin{multlined}
\label{eqn:f-r-sigma(b+v)}
 \hat{\sigma}_{\rm {B+V}} 
        =\hat{\sigma}_{\rm B} \Bigg[1+\frac{\alpha_s C_\epsilon}{2\pi}\Big\{C_F\Big[-\frac{2}{\epsilon_{\rm IR}^2}-\frac{3}{\epsilon_{\rm IR}}+\pi^2-8\Big]\\
                                      +C_D\Big[ \frac{11}{3}\frac{1}{\epsilon_{\rm UV}}-\frac{1}{\epsilon_{\rm IR}}-\frac{2\pi^2}{3}+\frac{77}{18}\Big] \Big\} \Bigg].
\end{multlined}
\end{equation}

To get rid of the UV divergence in the above, renormalization of the 
coupling parameter $\lambda$ is necessary which is equivalent to adding 
an UV counter term of the following form to 
Eq.~\ref{eqn:f-r-sigma(b+v)},
\begin{equation}
\label{eqn:UV-counter}
\hat{\sigma}_{\rm C.T.}^{\rm UV} = - \hat{\sigma}_{\rm B}\frac{\alpha_s}{2\pi}\frac{(4\pi)^\epsilon}{\Gamma(1-\epsilon)}C_D
\frac{11}{3}\Big(\frac{1}{\epsilon_{\rm UV}}+{\rm ln}\frac{\mu^2}{\mu_R^2}\Big)
\end{equation}
where $\mu_R$ is the renormalization scale.
Hence the UV renormalized parton-level cross section to ${\cal 
O}(\alpha_s)$ for the production of diquark from $qq$ initial state is 
given by 
\begin{eqnarray}
 \hat{\sigma}_{\rm B+V+C.T.} 
                         &=& \hat{\sigma}_{\rm B}  \Big[1 + \frac{\alpha_s C_\epsilon}{2\pi} 
                         \Big\{ C_F\Big(-\frac{2}{\epsilon_{\rm IR}^2} -\frac{3}{\epsilon_{\rm IR}} +\pi^2 -8 \Big)\nn \\
                         && +  C_D \Big(-\frac{1}{\epsilon_{\rm IR}} + \frac{11}{3}{\rm ln}\frac{\mu_R^2}{\hat{s}}  -\frac{2\pi^2}{3}+ \frac{77}{18}\Big) \Big\}\Big].
\label{virt_prod}
\end{eqnarray}
Note that the procedure of renormalization has introduced a scale dependence in 
the cross section which would help in reducing the overall scale dependence due to the running of 
the coupling. After regulating the UV divergence, we are left with IR divergences, part of which will be 
canceled (due to Kinoshita-Lee-Nauenberg (KLN) theorem \cite{muta}) once we take into account the 
real gluon emission contribution. It is important to note that the singularity structure of 
virtual cross section is the same in the scalar \cite{Han:2009ya} and vector diquark cases. Just like 
the singular terms proportional to $C_F$, we find that the singular term proportional to $C_D$ is also universal.

 \begin{figure}[ht]
 \begin{center}
 \includegraphics[width=0.7\linewidth,height=0.55\linewidth]{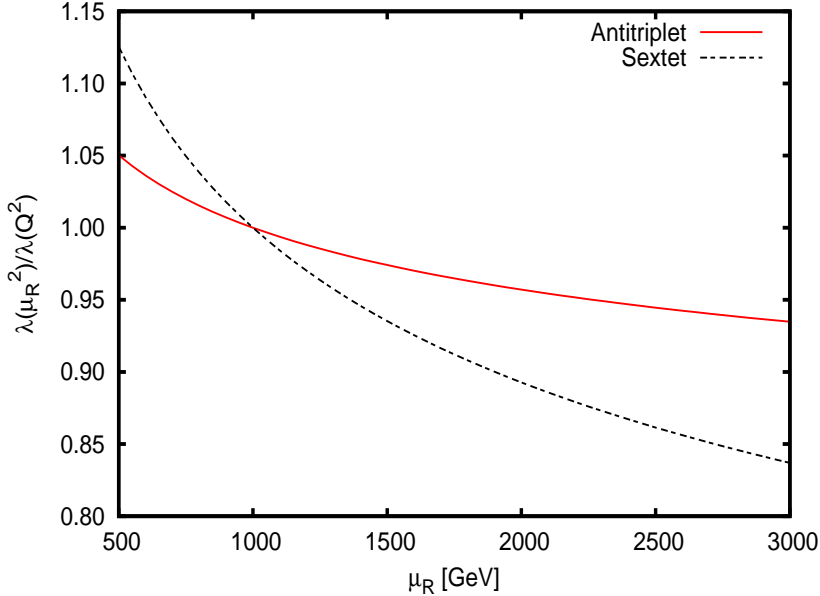}
 \end{center}
 \caption{Running of quark-diquark coupling with respect to the renormalization scale ($\mu_R$) to ${\cal O}(\alpha_s)$.}
 \label{fig:running-coupling-qqD}
 \end{figure}
Note that the results of this section can be utilized to predict the one-loop running of the quark-quark-diquark coupling 
$\lambda$. The one-loop beta function due to ${\cal O}(\alpha_s)$ QCD correction is therefore given by 
\begin{equation}
 \beta(\lambda) = \mu^2 \frac{d {\rm ln}\lambda}{d\mu^2} = -\frac{\alpha_s}{4\pi} \Big(\frac{11}{3}C_D\Big).
\end{equation}
Solving this, the running of the renormalized coupling parameter $\lambda(\mu_R^2)$ follows\footnote{ We 
would like to point out that in the expression of running coupling for the 
scalar diquark case,  given in Eq. 4.4 of Ref.~\cite{Han:2009ya}, the factor of $C_F$ should also  
be multiplied in the ${\cal O}(\alpha_s)$ term.} 
\begin{equation}\label{eq:running coupling}
 \lambda(\mu_R^2)=\lambda(Q^2)\Big[1-\frac{\alpha_s(\mu_R^2)}{4\pi}\frac{11}{3}C_D\ln\Big(\frac{\mu_R^2}{Q^2}\Big)\Big],
\end{equation}
where $Q$ is a reference scale which we will identify with $M_D$ (mass of the vector diquark) and  choose $\lambda(M_D^2) = 1$. 
It is worth pointing out that in contrast to the scalar diquark case, the one-loop running of the coupling in the vector 
diquark case depends on the diquark representation and therefore will behave differently for the antitriplet and the sextet.
This is highlighted in Fig. \ref{fig:running-coupling-qqD} where we show how the coupling $\lambda$ varies as a function of the
renormalization scale $\mu_R$. Note that we have chosen $\lambda(Q^2)=1$ for $Q=M_D=1$ TeV as a reference point which is just 
for illustration purposes only.  The scale dependence for the
antitriplet vector diquark coupling is found to be at $\sim 6\%$ for the $\mu_R$ range considered while that for the sextet turns out
to be significantly higher at $\sim 16\%$ for the same variation in $\mu_R$. This is due to the dependence of the one-loop beta function 
on $C_D$ which takes different values for the two cases.  Note that the running of the coupling will bring in a scale dependence 
for the LO cross section of the diquark too, similar to that observed for QCD cross sections due to the running of the strong 
coupling constant $\alpha_s$.

 \subsection{Real Corrections: $qq$ channel}
Next, we compute the contribution from the gluon bremsstrahlung 
radiated from initial state as well as final state to ${\cal O}(\alpha_s)$. 
The process for the real gluon emission is,
 \begin{equation}
  q_i (p_1) + q_j (p_2) \rightarrow g(k) + V_D (p_1+p_2-k).  \nn
 \end{equation}
 \begin{figure}[ht]
 \label{fig:qqd-real-nlo-feyn}
 \begin{center}
 \includegraphics[width=0.75\linewidth,height=0.25\linewidth]{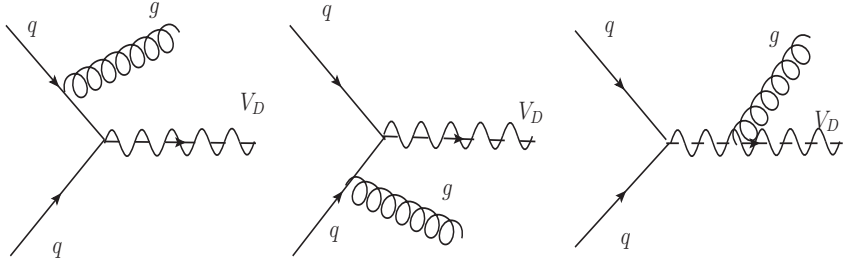}
 \end{center}
 \caption{Feynman diagram at leading order  for the process $qq\rightarrow g~ V_D .$}
 \label{fig:qqd-real-nlo-feyn}
 \end{figure}
The Feynman diagrams which contribute to the NLO level gluon emission 
process for diquark production  is given in 
Fig~\ref{fig:qqd-real-nlo-feyn}. The full ${\cal O}(\alpha_s) $ spin and color averaged 
squared-amplitude for the three different diagrams shown in Fig. \ref{fig:qqd-real-nlo-feyn} 
can be expressed in terms of Mandelstam variables $(s,t,u)$ in $n=4-2\epsilon$ space-time 
dimensions and is given by,
\begin{eqnarray}
 \sum \overline{|{\cal M}^R_{qq}|^2} &=& \frac{(n-2)N_Dg_s^2\lambda^2 \mu^{4\epsilon}}{4N_C^2tu(t+u)^2}
  \Big(-C_Dtu+C_F(t+u)^2\Big) \nonumber \\  &\times& \Big(4s^2+(n-2)t^2 +2(n-4)tu+(n-2)u^2+4s(t+u)\Big) 
\end{eqnarray}
where $s = (p_1+p_2)^2,\; t = (p_1-k)^2\;$ and $u = (p_2-k)^2$. 
The partonic cross section for the real gluon emission process is obtained by performing the phase space integration 
in $4-2\epsilon$ dimensions and is given  by
\begin{equation}
 \begin{multlined}
  \hat{\sigma}^{R}_{qq} = \frac{\hat{\sigma}_0}{\hat{s}} \frac{\alpha_s}{2\pi}C_\epsilon \Bigg[ C_F\Bigg\{(\frac{2}{\epsilon_{\rm IR}^2}
  +\frac{3}{\epsilon_{\rm IR}}-\frac{\pi^2}{3})\delta(1-\tau)-\frac{2}{\epsilon_{\rm IR}}\left(\frac{1+\tau^2}{1-\tau}\right)_{+} \\
  \shoveleft[2cm]
     +4(1+\tau^2)\left(\frac{{\rm ln}(1-\tau)}{1-\tau}\right)_{+}\Bigg\}\\
      + C_D
    \Bigg\{(\frac{1}{\epsilon_{\rm IR}}+\frac{11}{3}) \delta(1-\tau) -\frac{2}{3}\left(\frac{1+\tau+\tau^2}{1-\tau}\right)_{+}
   \Bigg\}+
    \mathcal{O}(\epsilon_{\rm IR})\Bigg]
\end{multlined}
\label{eq:sigma_qqR}
\end{equation}
In the above expression, the terms with $(\ldots)_+$  are the \textit{plus functions}. 
The plus function distribution is defined in Appendix \ref{app:PF}. The IR 
divergence of real emission process originates from the phase space 
region where the emitted gluon is soft ($k_0 \to 0$) and/or it is 
collinear to the quarks. Since $\tau=1$ corresponds to threshold 
production of the vector diquark, the $1/\epsilon_{\rm IR}$ singular terms 
proportional to $\delta(1-\tau)$ are due to the gluon becoming soft. On the 
other hand, the $1/\epsilon^2_{\rm IR}$ term arises when this soft 
gluon is also collinear to any of the two initial state quarks. The remaining 
singular terms in Eq. \ref{eq:sigma_qqR} are due to the gluon becoming 
collinear to quarks. Since the vector diquark is massive, the gluon emitted 
from it cannot be collinear thus explaining the absence of collinear singularity 
in $C_D$ part of the expression.  As mentioned above, the IR soft singularities 
cancel between real and virtual correction to $qq \to V_D$. Adding the two cross sections 
given by Eqs. \ref{eqn:f-r-sigma(b+v)} and \ref{eq:sigma_qqR}, we get
\begin{equation}
\begin{multlined}
 \hat{\sigma}_{\rm B+V+C.T.+R} = \hat{\sigma}_{\rm B+V+C.T.} + \hat{\sigma}^{\rm R}_{qq}\\
 \shoveleft[2.1cm]
                              = \frac{\hat{\sigma}_0}{\hat{s}}\Bigg[\delta(1-\tau)  
                              +\frac{\alpha_s C_\epsilon}{2\pi}\Bigg\{C_F\Big[\Big(\frac{2\pi^2}{3}-8\Big)\delta(1-\tau)
                              -\frac{2}{\epsilon_{\rm IR}}\left(\frac{1+\tau^2}{1-\tau}\right)_{+}\\ 
 \shoveleft[2.5cm]
                          + 4(1+\tau^2)\left(\frac{{\rm ln}(1-\tau)}{1-\tau}\right)_{+}\Big] \\
\shoveleft[2.2cm]                          
                          +C_D\Big[\Big(\frac{11}{3}{\rm ln}\frac{\mu_R^2}{\hat{s}}-\frac{2\pi^2}{3}+\frac{143}{18}\Big)\delta(1-\tau)
                          -\frac{2}{3}\left(\frac{1+\tau+\tau^2}{1-\tau}\right)_{+}\Bigg\}\Bigg],
\end{multlined}
\end{equation}
where we are left only with the collinear divergence terms as expected. The 
collinear divergences can be finally removed by redefining the quark PDF's.  
In the $\overline{\rm MS}$ factorization scheme, the universal counter term for collinear singularity is 
\begin{equation}
 \hat{\sigma}^{\rm C.T.}_{qq} =   \frac{\hat{\sigma}_0}{\hat{s}} \frac{\alpha_s}{2\pi} 
   \frac{(4\pi)^\epsilon}{\Gamma(1-\epsilon)}
   \Big(\frac{1}{\epsilon_{\rm IR}}+{\rm ln}\frac{\mu^2}{\mu_F^2}\Big) 2 P_{qq}(\tau)
\end{equation}
where
$
P_{qq}(\tau) = C_F\Bigg(\frac{1+\tau^2}{1-\tau}\Bigg)_{+}
$
is the Dokshitzer-Gribov-Lipatov-Altarelli-Parisi (DGLAP) splitting function (probability of quark splitting into a 
quark and a gluon) and $\mu_F$ defines the factorization scale. The total parton level cross section in $qq$ channel is finally given by,
\begin{equation}
 \begin{multlined}
\hat{\sigma}_{qq} =\hat{\sigma}_{\rm B+V+C.T+R}+\hat{\sigma}^{\rm C.T.}_{qq}\\
\shoveleft[0.2cm]   =
\frac{\hat{\sigma}_{0} }{\hat{s}}\Bigg[\delta(1-\tau) + \frac{\alpha_s}{2\pi}\Bigg\{
  2P_{qq}(\tau){\rm ln}\Big(\frac{M_D^2}{\mu_F^2\tau}\Big)
  + C_F\Big[4(1+\tau^2)\Bigg(\frac{{\rm ln}(1-\tau)}{1-\tau}\Bigg)_{+}
 + (\frac{2\pi^2}{3}-8)\delta(1-\tau)\Big]
 \\ \shoveleft[2.4cm] 
  + C_D\Big[-\frac{2}{3}\Bigg(\frac{1+\tau+\tau^2}{1-\tau}\Bigg)_{+} + 
  \Big(\frac{11}{3}{\rm ln}\Big(\frac{\mu_R^2}{M_D^2}\Big)-\frac{2\pi^2}{3}
  +\frac{143}{18}\Big)\delta(1-\tau)\Big]\Bigg\}\Bigg].
 \end{multlined}
 \label{eqn:qq-NLO}
\end{equation}
The corresponding hadronic cross section is obtained by convoluting the 
parton level cross section with the initial state quark distribution functions, 
\begin{equation}
  \sigma_{qq} = \int_{\tau_0}^1 d\tau~ \frac{\tau_0}{\tau^2}~\Big[(q\otimes q)\Big(\frac{\tau_0}{\tau}\Big)\Big]~{\hat \sigma}_{qq}.
\end{equation}
If the initial state quarks are of different flavors $q_1$ and $q_2$ then replace, $q\otimes q \to (q_1\otimes q_2 + q_2\otimes q_1)$ 
in the above equation. 

\subsection{Real Corrections: $qg$ channel}
As pointed out earlier, for a complete ${\cal O}(\alpha_s)$ contribution we should 
also consider the quark-gluon ($qg$) initiated process,
 \begin{equation}
  q_i (p_1) + g(k) \rightarrow V_D (p_1+k-p_2) + \overline{q}_j (p_2). \nn
 \end{equation}
 \begin{figure}[ht]
 \begin{center}
 \includegraphics[width=0.85\linewidth,height=0.25\linewidth]{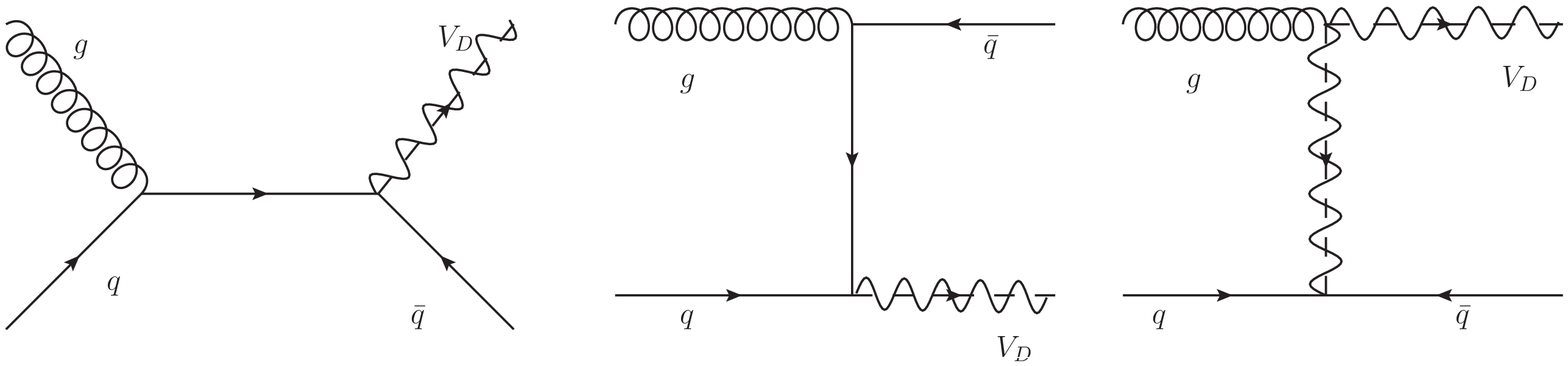}
 \end{center}
 \caption{Feynman diagrams  for the process $qg\rightarrow \bar{q} ~V_D.$}
 \label{fig:qqd-gind-nlo-feyn}
 \end{figure}
 
The Feynman diagrams for this process are given in 
Fig~\ref{fig:qqd-gind-nlo-feyn}. The total spin and color averaged amplitude-squared for the 
$qg$ initiated process in terms of Mandelstam variables  is given by 
\begin{equation}
  \begin{multlined}
  \sum \overline{|\mathcal{M}_{qg}^R|^2} = \frac{g_s^2\lambda^2 \mu^{4\epsilon} N_D (C_Dsu-C_F(s+u)^2)}{2 N_C (N_C^2-1)su(s+u)^2}
  \\ \shoveleft[2cm] \times
  \Big ((n-2)(s^2+u^2)+4t(u+s)+2(n-4)su+4t^2 \Big)
  \end{multlined}
  \end{equation}
Note that the spin average for the initial state gluon introduces a term dependent 
on the space-time dimension $n$ and also has a different color averaging factor compared to the 
$qq$ initiated process for real corrections. However, as expected the above expression  does match 
with that for the $qq$ case without the spin and color averaging, under the  interchange 
$t \leftrightarrow s$ and an overall sign. This is because of the crossing symmetry between $qq$ and $qg$ processes. 
The extra -ve sign in $qg$ case results when  one fermion is moved from initial state to the final state. 

The parton level cross section for the $qg$ initiated process is 
\begin{equation}
 \begin{multlined}
  \hat{\sigma}_{qg}^{R} = \frac{\hat{\sigma}_0}{\hat{s}}\frac{\alpha_s}{2\pi} C_\epsilon
   \Bigg[\Bigg\{\frac{-((1-\tau)^2+\tau^2)}{2\epsilon_{\rm IR}} + \frac{3+2\tau-3\tau^2}{4} 
+((1-\tau)^2+\tau^2){\rm ln}(1-\tau)\Bigg\}\\
+ \frac{C_D}{2C_F}\Bigg\{-1+\frac{2}{\tau}+\tau-2\tau^2+2(1+\tau){\rm ln}\tau\Bigg\}\Bigg],
 \end{multlined}
\end{equation}
where $\hat\sigma_{0}$ is given in Eq.~\ref{eqn:sigma_0}.
As shown above, the cross section has IR collinear divergence which we remove by 
factorization in $\overline{\rm MS}$ scheme. The required counter term is given by,
\begin{equation}
 \sigma_{qg}^{\rm C.T.} = \frac{\hat{\sigma}_0}{\hat{s}}\frac{\alpha_s}{2\pi} 
\frac{(4\pi)^\epsilon}{\Gamma(1-\epsilon)}
   \Big(\frac{1}{\epsilon_{\rm IR}}+{\rm ln}\frac{\mu^2}{\mu_F^2}\Big)  P_{qg}(\tau)
\end{equation}\\
with $ P_{qg}(\tau) = \frac{1}{2}\Big[(1-\tau)^2+\tau^2\Big] $.
Hence the parton level cross section for the vector diquark production in $qg$ initiated channel is given by
\begin{equation}
 \begin{multlined}
  \hat{\sigma}_{qg}=\hat{\sigma}_{qg}^{R}+\sigma_{qg}^{\rm C.T.}\\
  \shoveleft[0.6cm]  =
  \frac{\hat{\sigma}_0}{\hat{s}}\frac{\alpha_s}{2\pi}\Bigg[P_{qg}(\tau)
  \Big\{{\rm ln}\Big(\frac{M_D^2}{\mu_F^2\tau}\Big) + 2{\rm ln}(1-\tau)\Big\} +\frac{3+2\tau-3\tau^2}{4} \\
  \shoveleft[1.2cm] +\frac{C_D}{2C_F}\Big\{-1+\frac{2}{\tau}+\tau-2\tau^2
  +2(1+\tau){\rm ln}\tau\Big\}\Bigg].
 \end{multlined}
 \label{eqn:qg-NLO}
\end{equation}
The corresponding hadronic cross section is obtained by convoluting the 
above parton level cross section with the initial state quark and gluon 
distribution functions,
\begin{equation}
 \sigma_{qg} = \int_{\tau_0}^1 d\tau~ \frac{\tau_0}{\tau^2}~\Big[(q\otimes g + g\otimes q)\Big(\frac{\tau_0}{\tau}\Big)\Big]~{\hat \sigma}_{qg}.
\end{equation}

\section{Decay Width: ${\cal O}(\alpha_s)$ correction} 
\label{sec:decay}
Note that just like the LO cross sections for the production of the vector diquarks, the LO predictions for decay width 
of the particle also suffer from the renormalization scale uncertainties. Therefore for the sake of 
completion we would also like to estimate the effect of the QCD corrections on the decay width ($\Gamma$) of the vector diquark.  
Note that a primary requirement in assuming the narrow width approximation, one expects that the ratio $\Gamma/M_D$  is relatively small 
and not exceeding $\simeq 10\%$.  In order to remain in that regime, it is necessary to check that the decay width does not change by much  
under higher-order corrections.  In this section,  we compute the NLO QCD corrections to the vector diquark decaying into a 
pair of light jets,
\beq
V_D(q) \rightarrow q_i(p_1) + q_j(p_2).
\label{LO_dk_process}
\eeq
The leading order total decay width is given by
\beq
\Gamma_0 = \sum_{i}~{\lambda_i^2 \over 24 \pi} M_D,
\eeq
where $i$ is the number of light quark generations which can couple to the vector diquark of a given electric charge. We have
assumed that we can neglect all quark masses in the decay products (including top quark).
The virtual corrections to the decay width involve the same Feynman graphs shown in Fig. \ref{fig:nlo-virtual-feyn} and has the 
same singular structure as given in Eq. \ref{virt_prod} for the on-shell vector diquark production. The same procedure followed in 
calculating the virtual corrections for the production cross section leads us to the UV renormalized virtual correction to the decay 
width which is given by
\barr
\Gamma^V &=& \Gamma_0 \Bigg[1 + {\alpha_s \over 2 \pi} C_{\epsilon}\Bigg\{C_F\Bigg(-{2\over \epsilon_{IR}^2} -{3\over \epsilon_{IR}} 
-8+\pi^2\Bigg) \nonumber\\[1ex]
&&+ C_D\Bigg(-{1\over \epsilon_{IR}} + {11\over 3}\ln\Big({\mu_R^2\over M_D^2}\Big) + {77\over 18} - {2 \pi^2\over 3}\Bigg)\Bigg\}\Bigg].
\label{virt_dk}
\earr
However we must point out that the real gluon correction is inherently different from that of the production. To compute the real 
gluon correction to the decay width, we need to consider the following three body final state, 
\barr
V_D(q) \rightarrow q_i(p_1) + q_j(p_2) + g.
\label{eq:dk_process_R}
\earr
Note that the calculation of real correction to diquark decay width requires three 
body phase space integration to be performed in $n=4-2\epsilon$ dimensions. For 
that we have followed the method given in Ref.~\cite{Field}. The final expression with the 
real correction to the decay width is then given by,
\barr
\Gamma^R &=& \Gamma_0 {\alpha_s \over 2 \pi} C_{\epsilon}\Bigg[C_F\Bigg({2\over \epsilon_{IR}^2} +{3\over \epsilon_{IR}}
+{19\over 2}-\pi^2\Bigg) 
+ C_D\Bigg({1\over \epsilon_{IR}} + {11\over 3}\Bigg)\Bigg].
\label{real_dk}
\earr
By adding the virtual and real corrections to decay width all the singularities cancel as expected by KLN theorem. Thus, 
the complete NLO QCD correction to the diquark decay width is given by (from Eq. \ref{virt_dk} and Eq. \ref{real_dk}),
\beq
\Gamma_{\rm NLO} = \Gamma_0 \Bigg[1 + {\alpha_s \over 2 \pi} \Bigg\{{3 \over 2}C_F + C_D\Bigg({143\over 18} - 
{2 \pi^2\over 3}+{11\over 3}\ln\Big({\mu_R^2\over M_D^2}\Big)\Bigg)\Bigg\}\Bigg]. 
\label{NLO_dk}
\eeq
The corresponding expression for the case of {\it scalar} diquark is given in Appendix~\ref{app:SDW-NLO}.
From Eq. \ref{NLO_dk}, we observe that the coefficient of $C_F$ is similar to that in SM (NLO QCD correction of an 
electroweak vector boson decaying into quark-antiquark pair) although here the final state is a quark-quark pair. We also find that a non-trivial 
contribution to the NLO decay width arises from the other Casimir, $C_D$ which takes different values for the two color 
representations of the vector diquark.

 \begin{figure}[t!]
 \includegraphics[width=0.5\linewidth,height=0.4\linewidth]{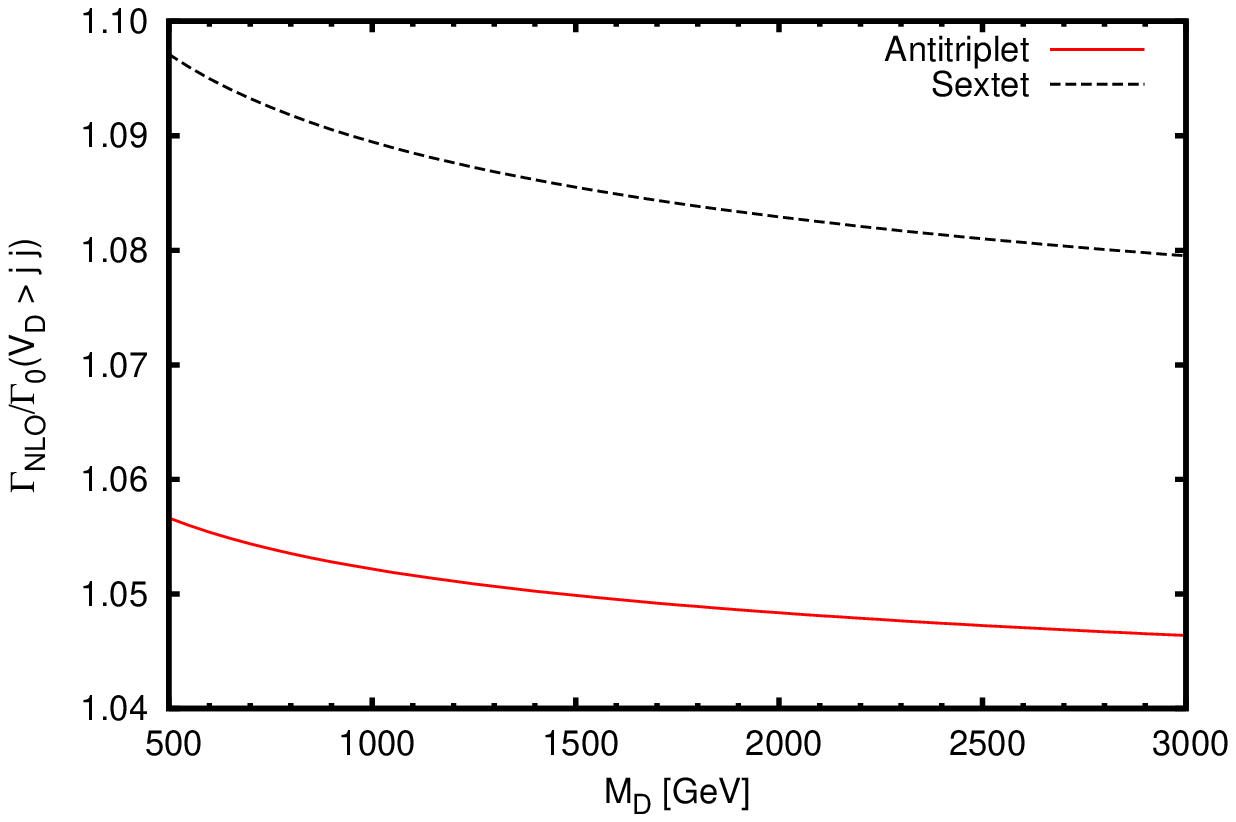}
 \includegraphics[width=0.5\linewidth,height=0.4\linewidth]{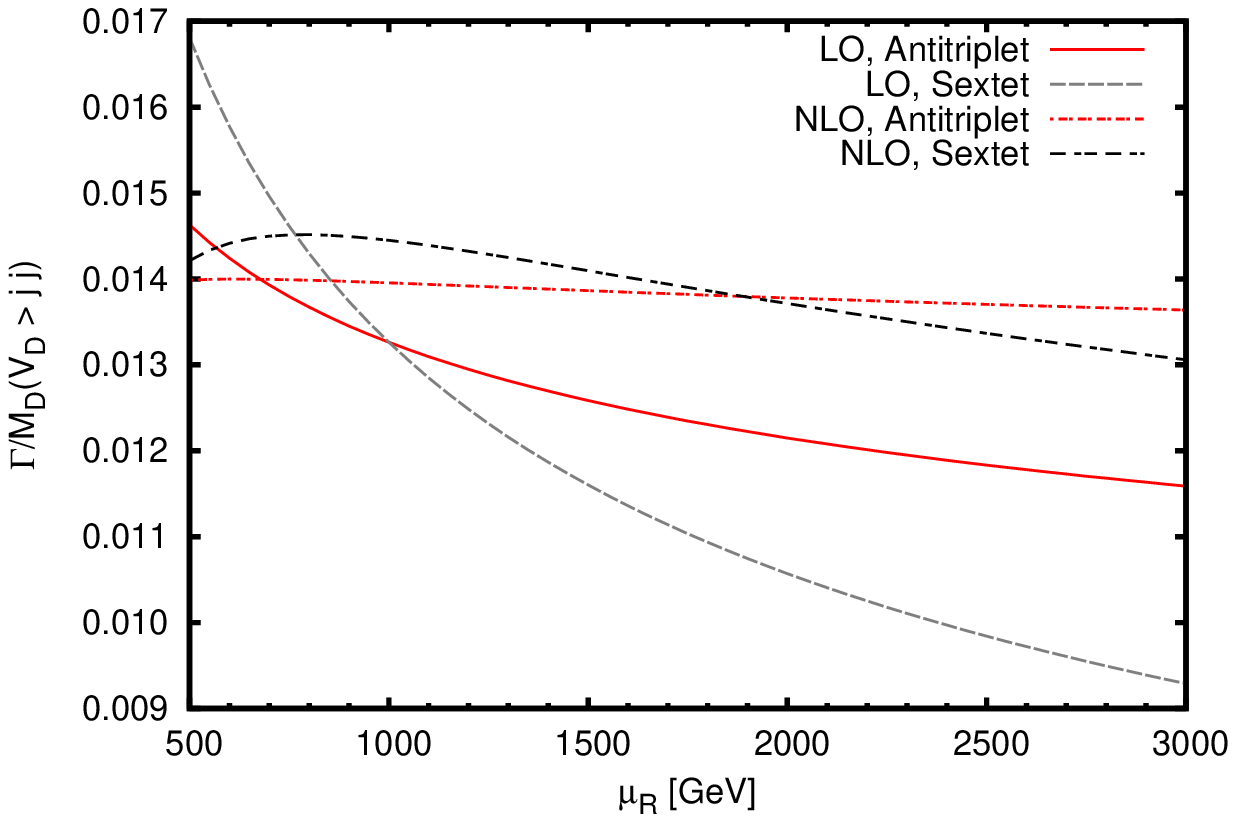}
 \caption{Dependence of decay width on $M_D$ (left) and on renormalization scale $\mu_R$ (right) to ${\cal O}(\alpha_s)$. 
 To show the $\mu_R$ dependence we chose $\lambda (M_D) =1$ where $M_D=1$ TeV.}
 \label{fig:width-nlo}
 \end{figure}
We calculate the relevant $K$-factor defined as the ratio of the NLO width to that of the LO width and plot it in Fig.  \ref{fig:width-nlo}. 
In the left panel of Fig.~\ref{fig:width-nlo}, we have shown the dependence of the NLO $K-$factor for the 
decay width on the diquark mass. As $\mu_R=M_D$ we can clearly see that the logarithmic term in Eq. \ref{NLO_dk} will not contribute
and we should expect a constant value for a particular diquark representation.  We however observe  a slight variation for the NLO $K$-factor 
for the widths of the antitriplet and sextet vector diquarks as we vary the mass $M_D$, which is only arising because of the running of the 
strong coupling $\alpha_s$ (we have taken $\alpha_s(M_Z)=0.1184$
as the reference value).  We find that $K$-factor for the sextet case is larger than the antitriplet due to larger $C_D$ and 
increases the LO width by about $8-10\%$ for the mass range $M_D=0.5-3$ TeV.  The corresponding LO width for the 
antitriplet vector diquark is modified less and increases by about $4.5-6\%$ with the $K$-factor. 
On the right panel of Fig.~\ref{fig:width-nlo}, we show the scale dependence $\mu_R$ of the 
decay width at LO and NLO and for sextet and antitriplet vector diquark states. As a reference point, we have chosen 
$\lambda (M_D)=1$  where $M_D=1$ TeV and we vary $\mu_R$ between $M_D/2$ to $3M_D$. The LO scale dependence 
is entirely due to the running of the coupling (see Fig.~\ref{fig:running-coupling-qqD}). We can clearly see that the 
inclusion of ${\cal O}(\alpha_s)$ correction has significantly reduced the scale dependence. As one 
would expect, due to the smaller color factor $C_D$ the scale variation for the antitriplet case 
is also smaller as compared to the scale variation for the sextet case.

\begin{figure}[b!]
\includegraphics[width=0.5\linewidth,height=0.4\linewidth]{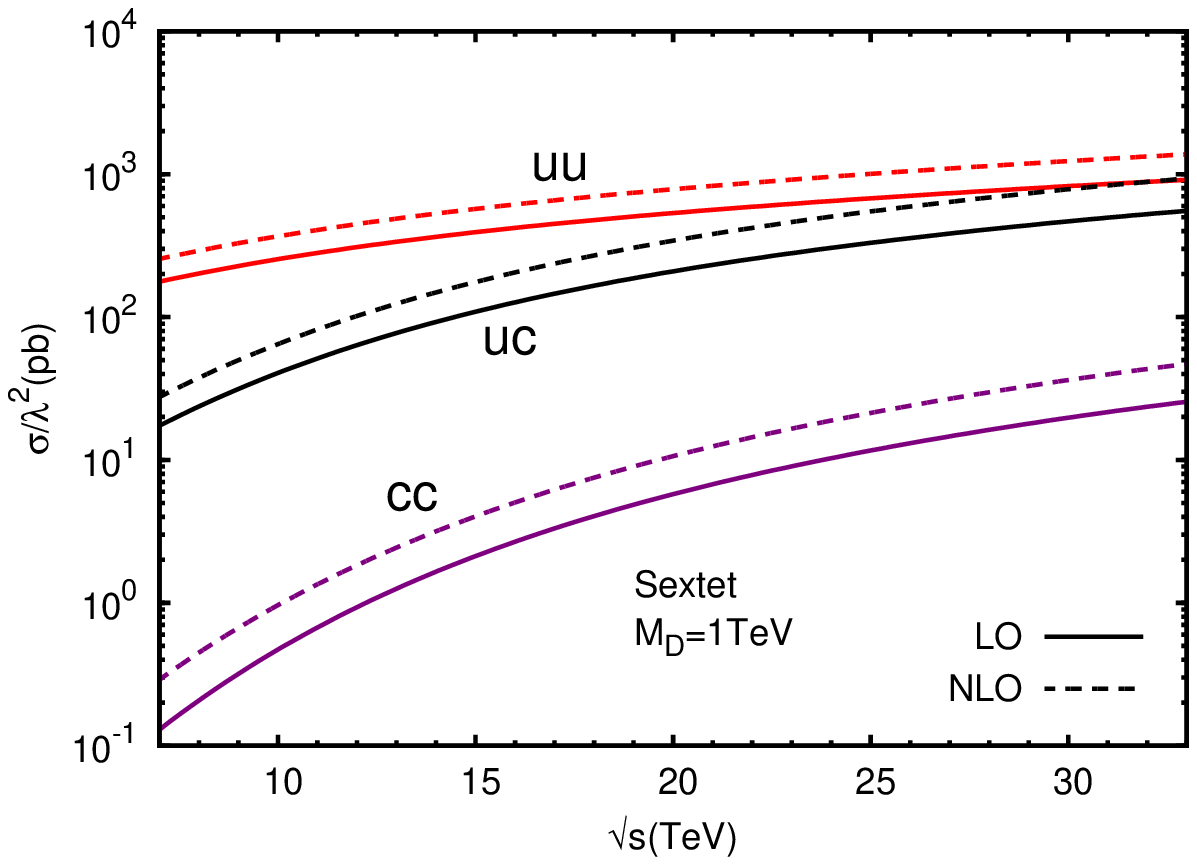}
\includegraphics[width=0.5\linewidth,height=0.4\linewidth]{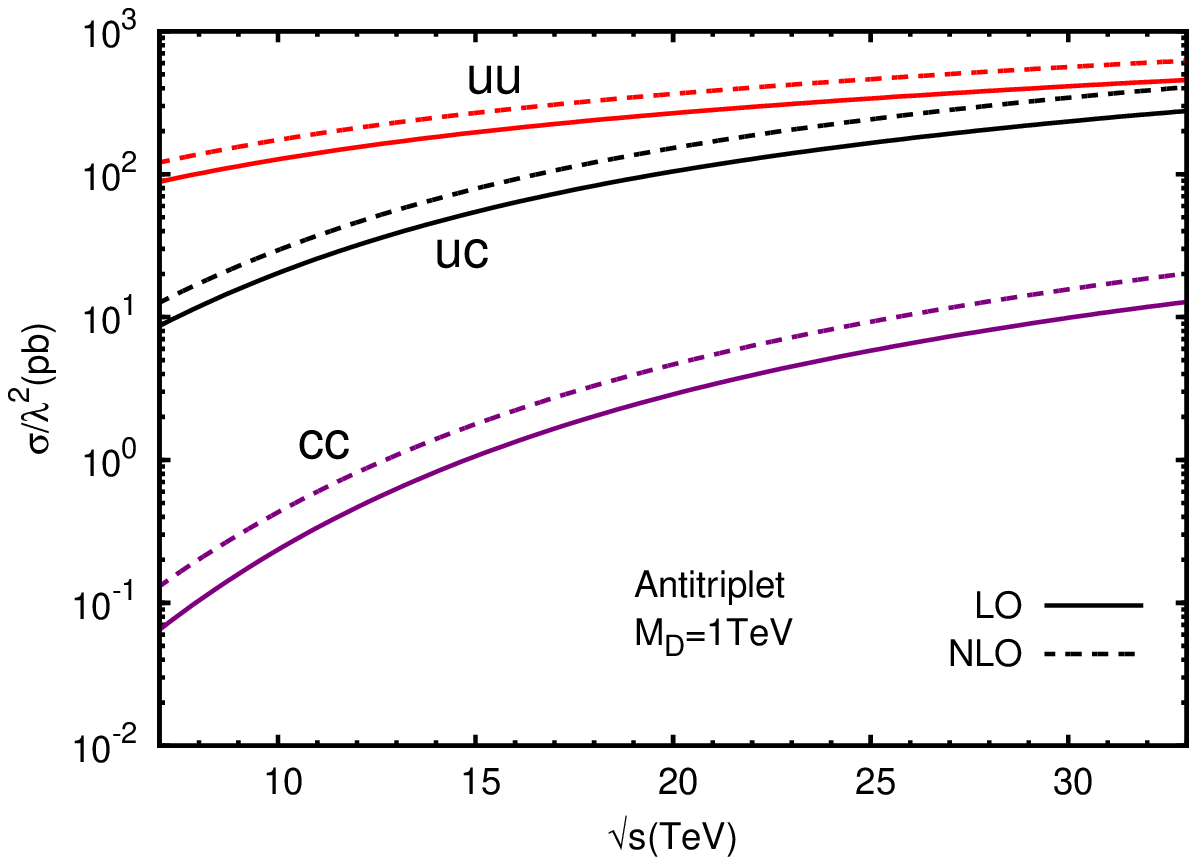}
\caption{Vector diquark production cross sections for the sextet and antitriplet cases at LO and NLO through the $uu,~uc$ and $cc$ initial states  
as a function of the $pp$ hadronic center of mass energy.}
\label{fig:cs_ecm}
\end{figure}
\section{Numerical Analysis and Results} \label{sec:numerics}
In this section, we discuss the LO and NLO results for the vector diquark production at the LHC. We have used the 
{\tt CTEQ6L1} ({\tt CTEQ6M}) \cite{Pumplin:2002vw} PDF's for the parton fluxes in the colliding protons for our LO (NLO) results. 
In our calculations we choose $\mu_F = \mu_R = M_D$ as the central scale for factorization and renormalization unless otherwise stated. 
Using our analytic results for the vector diquark production derived in the previous section we can now study how the cross sections 
are affected as a function of the collider center of mass energy ($\sqrt{s}$) as well as for different values of the mass ($M_D$) of the vector diquark.  
The LHC has already completed its run at two different $\sqrt{s}$ of $7$ and $8$ TeV and there are plans of running the machine at 
$13$ and $14$ TeV while future upgrades to $33$ TeV is also possible. In Fig.~\ref{fig:cs_ecm} we show  the LO and NLO 
hadronic cross sections for the on-shell vector diquark production as a function of  the proton-proton collider center of mass energy, for 
a fixed value of $M_D = 1$ TeV. Note that the variation observed in the LO cross section can be attributed to the initial parton PDF's 
only where, as the center of mass energy rises the on-shell condition of the diquark production for $M_D=1$ TeV forces the colliding 
partons to carry a much smaller $x$ (momentum fraction) of the proton beam energy. Therefore the initial quark's flux grows giving 
rise to increase in the production cross section.  The variation of the NLO cross section is however governed by both the 
partonic cross section and the PDF's although the feature attributed to the LO behavior due to the PDF's is similar.  
The plot  is shown for three different quark-quark initial states, namely $uu, cc$ and $uc$.
%
\begin{figure}[h]
\includegraphics[width=0.5\linewidth,height=0.4\linewidth]{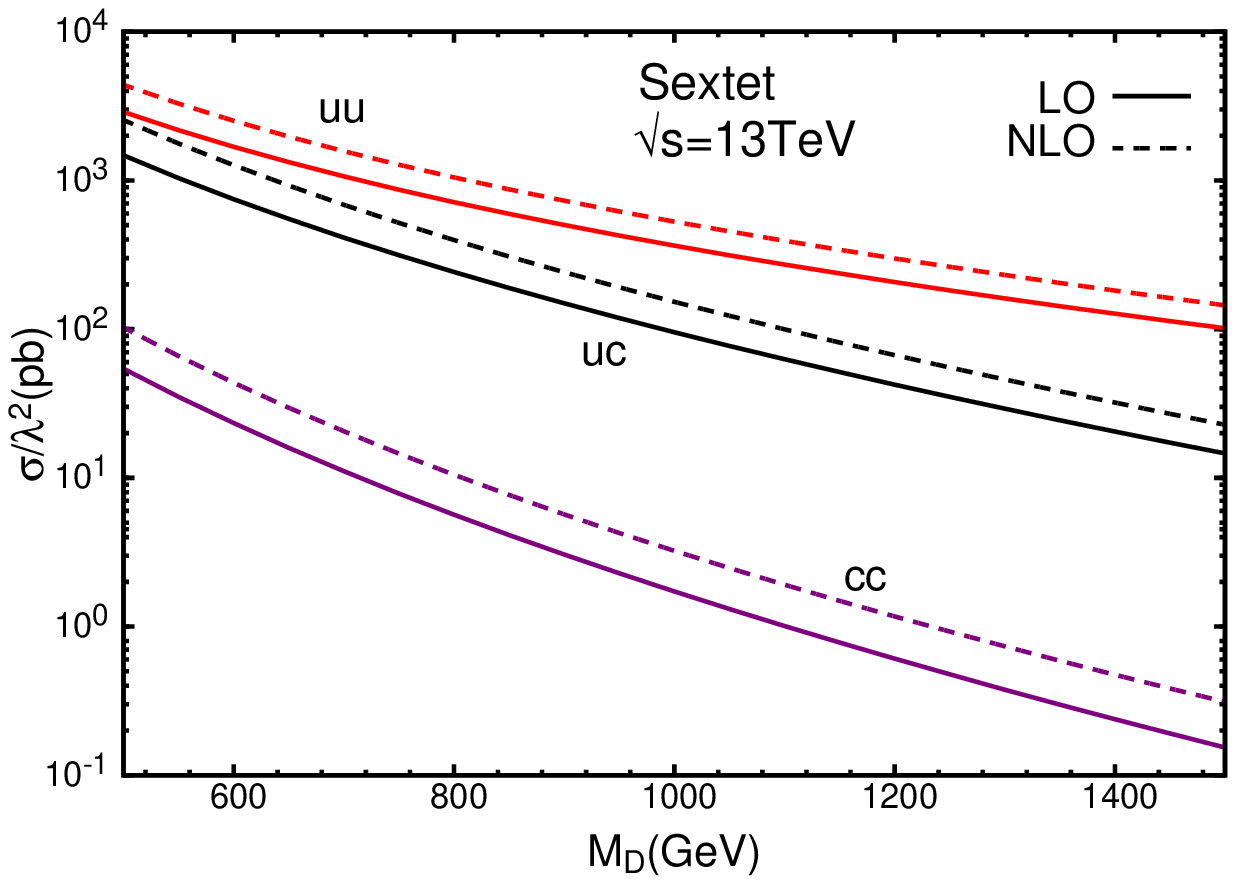}
\includegraphics[width=0.5\linewidth,height=0.4\linewidth]{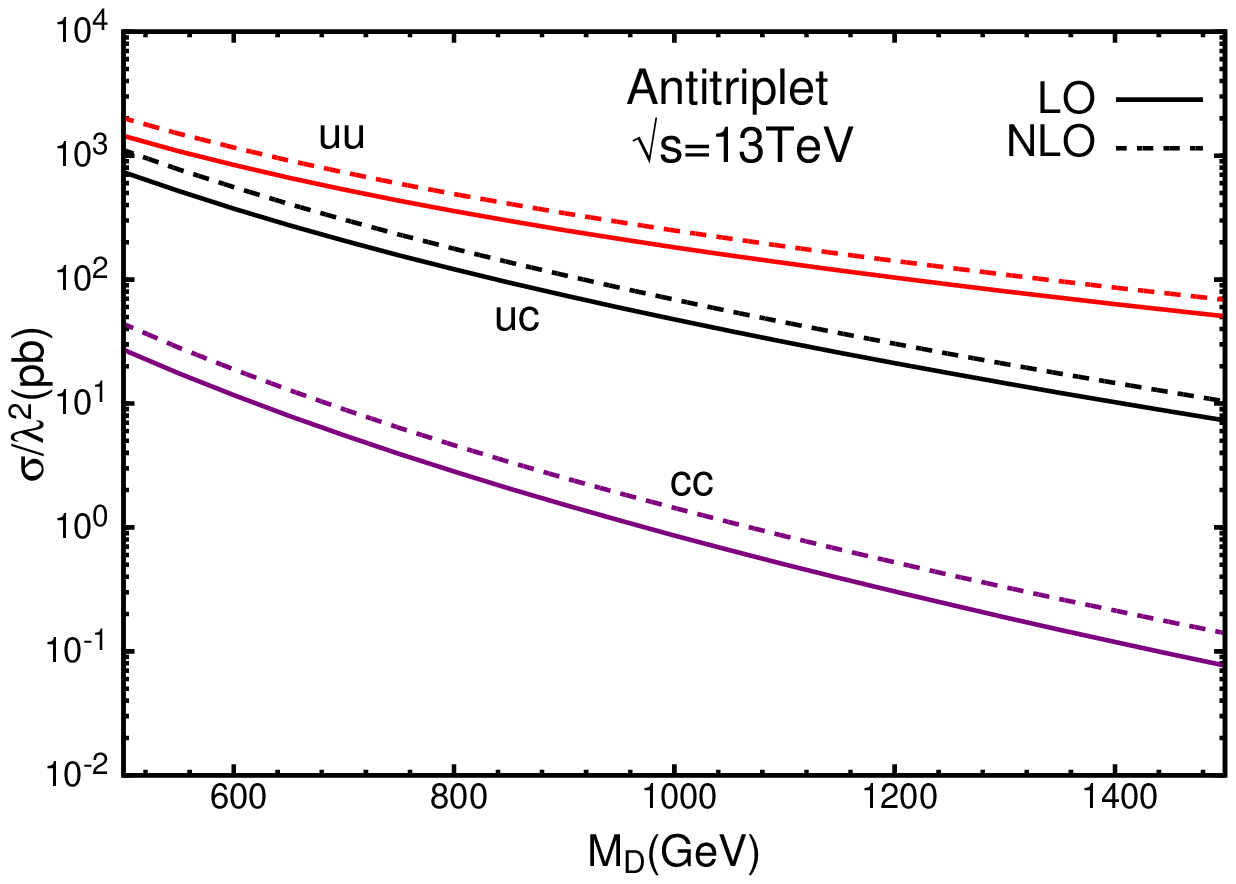}
\caption{Production cross section of the sextet and antitriplet vector diquarks at LO and NLO through the $uu, ~ uc$ and $cc$ initial states 
as a function of the diquark mass $M_D$ at LHC with $\sqrt{s}=13$ TeV.}
\label{fig:cross-sec-uu-cc-uc}
\end{figure}
It is worth recalling the fact that the coupling of the vector diquark can be generation and flavor dependent. 
Therefore one can consider the diquark  to be produced through initial partons of a particular fermion generation and flavor or it can 
be produced, mediated by interactions between different generations. We have chosen to normalize the cross sections with the coupling 
strength $\lambda$ squared so that it does not play a role here. Also note that although we always choose $\lambda (M_D)=1$ we have 
neglected the effect of the running of the coupling constant $\lambda$ in Fig. \ref{fig:cs_ecm}. Quite clearly, cross sections for the valence 
quark initiated processes are significantly large and reach appreciably high rates of above $\sim 100$ picobarns (pb) for ${\mathcal O(1)}$ coupling strengths. 
Even the {\it sea} quark rates rise from a few $100$ femtobarns (fb) to few ten's of picobarns for both the sextet and antitriplet vector diquarks 
for ${\mathcal O(1)}$ coupling strengths. When compared with the scalar diquark production rates we note that the LO cross section for the 
vector diquark production is exactly twice that of the scalar diquark.\footnote{In Ref. \cite{Han:2009ya} the interaction Lagrangian has an extra 
factor of $2\sqrt{2}$ thus giving overall rates higher than what we get here for the vector case. However once that is taken into consideration, one 
gets larger rates for the vector case as expected.}  Again, as against the scalar case where same flavor initial states are disallowed for the antitriplet 
case because of the antisymmetric property of the $K_{ab}$, one gets all modes contributing in the vector case \cite{Han:2010rf}. Thus a vector diquark 
which transforms as an antitriplet under $SU(3)_C$ would be produced through the initial valence $uu$ and $dd$ states resulting in a much higher 
cross section for the dijet final state compared to the scalar diquark which would have dominant production mode through $ud$ initial states.   
One important point to note here is that if only flavor diagonal couplings are allowed for the $uu$ type interactions then the vector antitriplet 
diquark will mediate same-sign top pair productions while the scalar diquarks will not, which would be a very interesting signal at the LHC.

\begin{figure}[h!]
\includegraphics[width=0.5\linewidth,height=0.4\linewidth]{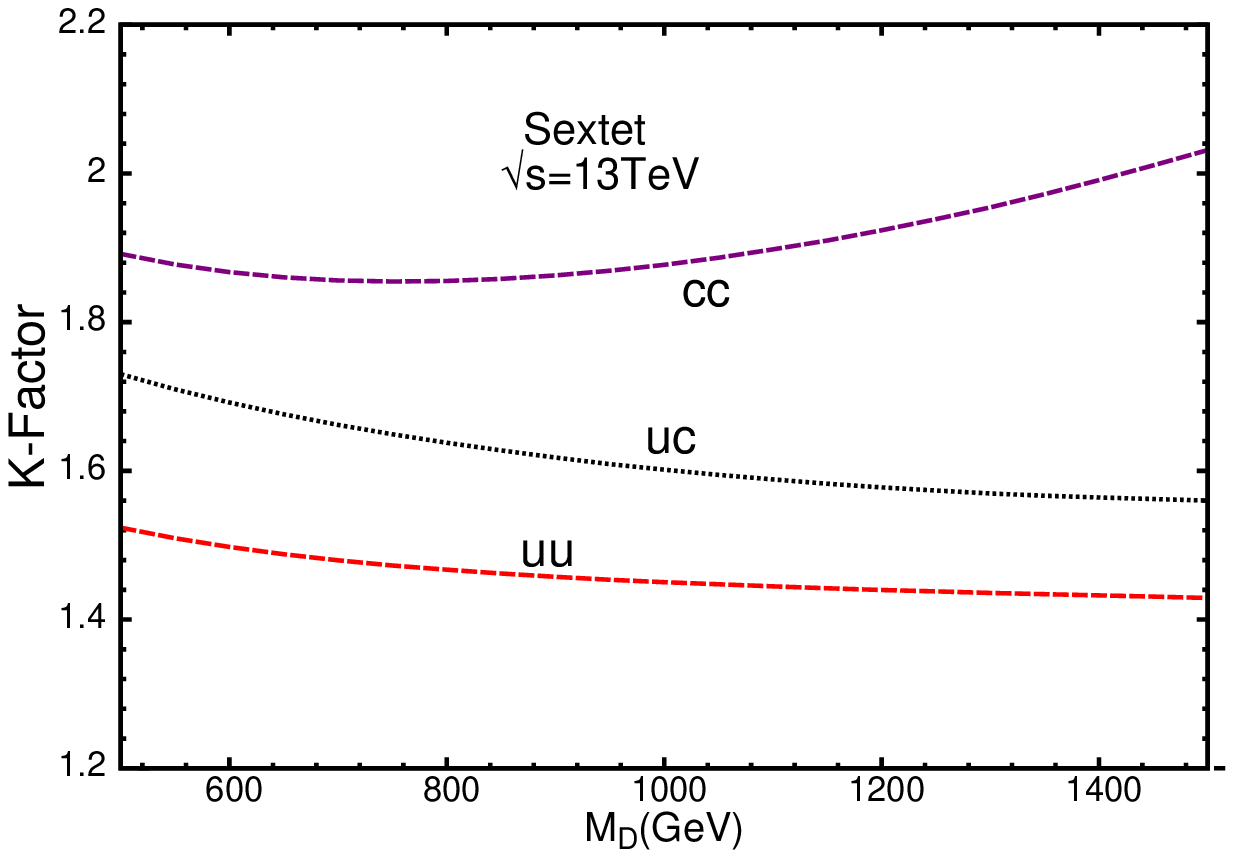}
\includegraphics[width=0.5\linewidth,height=0.4\linewidth]{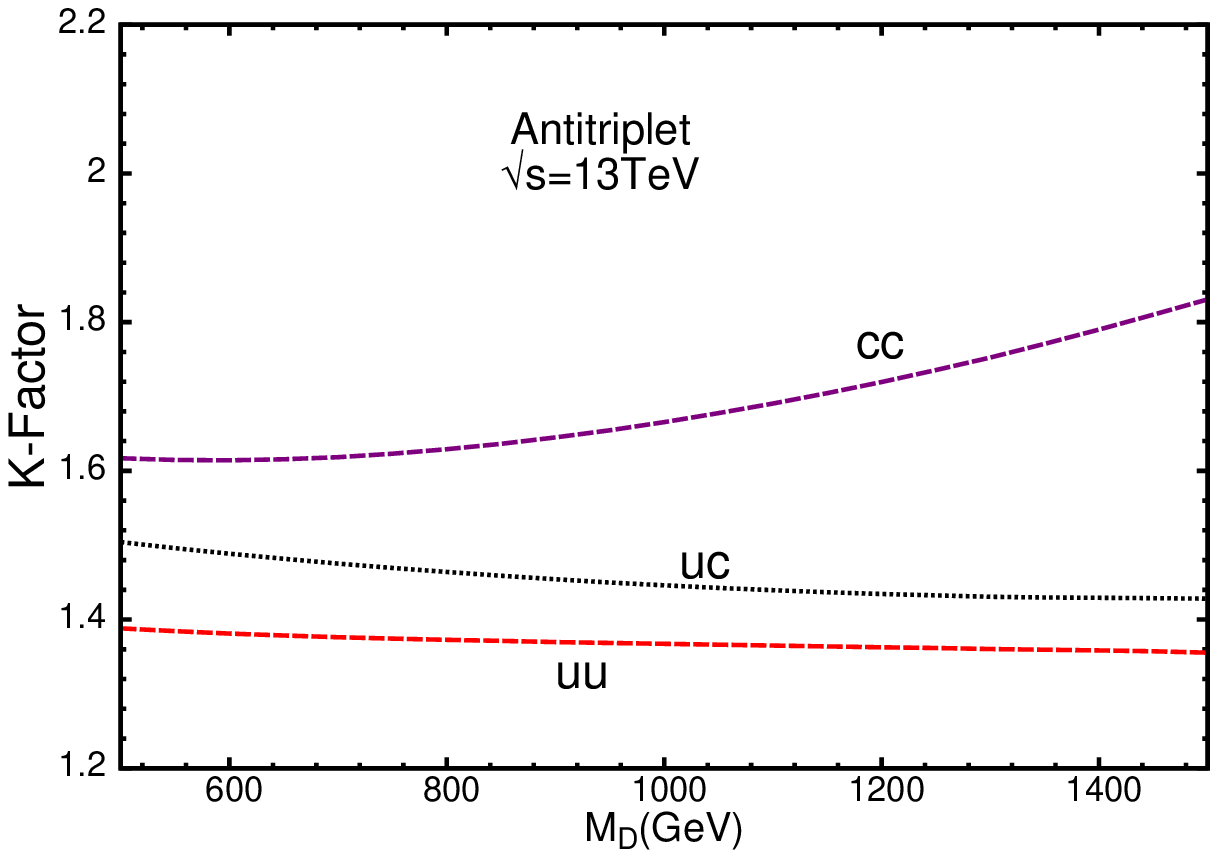}
\caption{Illustrating the NLO $K$-factors for the production of both sextet and antitriplet vector diquark at the LHC with $\sqrt{s}=13$ TeV,  
through the initial states $uu, ~uc$ and $cc$ as a function of the vector diquark mass $M_D$.}
\label{fig:k-factor-uu-cc-uc}
\end{figure}
Since the vector diquark mass ($M_D$) is a free parameter, it is also instructive to know how the production cross section varies as a function of the 
diquark mass.  We plot both the LO and NLO cross sections as a function of $M_D$ at the LHC run with $\sqrt{s}=13$ TeV in 
Fig.~\ref{fig:cross-sec-uu-cc-uc}.  
The plot  is again shown for three different  initial state combinations of quarks, namely $uu, cc$ and $uc$. All these would lead to the 
production of a vector diquark of charge +4/3.  The coupling strength has been factored out as before. 
We have varied $M_D$ in the range between $500$ GeV to $1.5$ TeV. Due to phase space suppression, the cross section 
goes down as we increase $M_D$. It is worth pointing it out here that due to the difference in $N_D$, the  sextet 
diquark production cross section at LO  is just twice that of the antitriplet production cross section (see Eq. \ref{eqn:sigma_0}). 
However, the NLO cross sections are markedly different for the two cases and therefore the NLO cross sections for the sextet
are no longer twice that of the antitriplet production. This will be evident from the $K$-factor estimates which we show next.    
Note that as all the different charged vector diquark productions are driven by the same color algebra for a given representation of $SU(3)_C$
the cross sections for them are eventually driven by the initial quark PDF's that participate in the production. Therefore the 
nature of the plots for the production cross section for the $|Q|=2/3,1/3$ charged diquarks  is very similar.

\begin{figure}[ht]
\includegraphics[width=0.5\linewidth,height=0.85\linewidth]{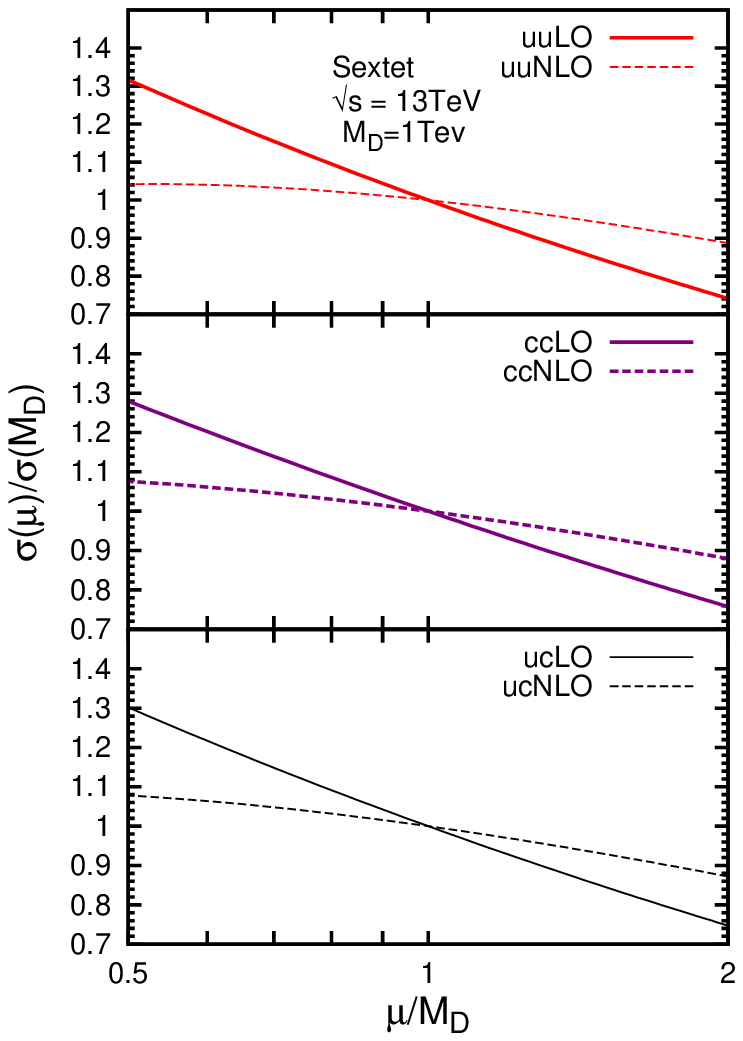}
\includegraphics[width=0.5\linewidth,height=0.85\linewidth]{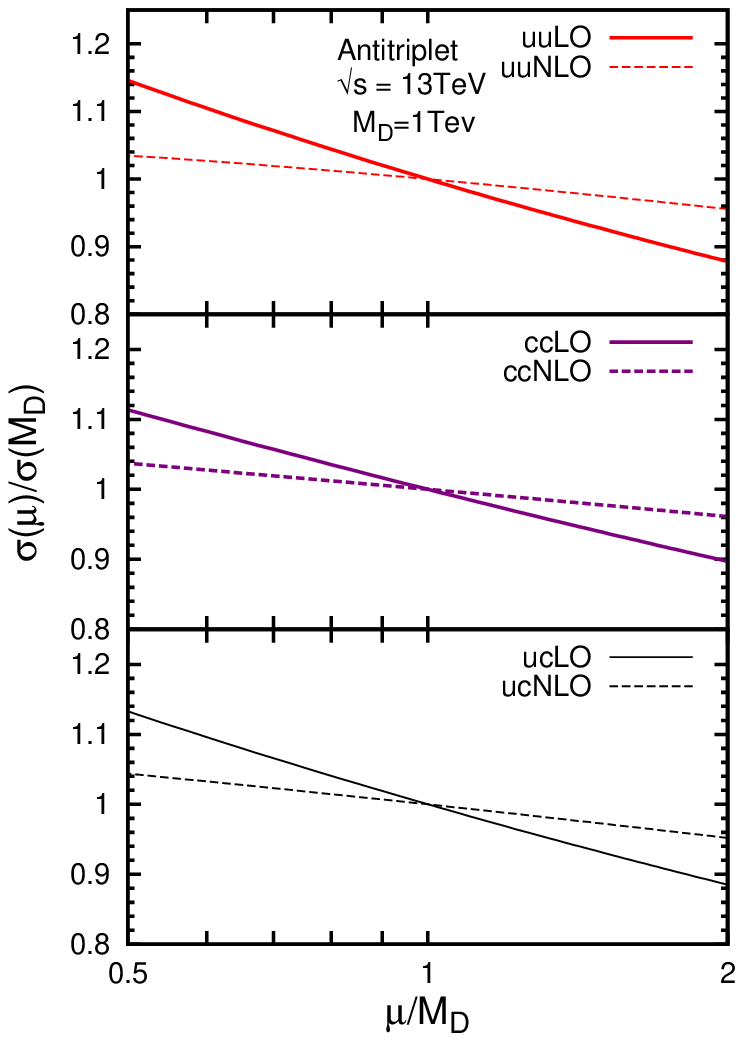}
\caption{Showing the scale dependence of LO and NLO production cross sections for sextet and
antitriplet diquark states of mass $M_D=1$ TeV at the LHC with $\sqrt{s} = 13$ TeV .}
\label{fig:13tev-uu-uc-cc-scale2}
\end{figure}
In Fig.~\ref{fig:k-factor-uu-cc-uc} we show the dependence of NLO $K$-factor, defined as the ratio of the 
NLO cross section to the LO cross section, on the vector diquark mass $M_D$ for both sextet and antitriplet
diquark states. The $K$-factors for the $uu$ and $dd$ initiated production are between 1.5 and 1.3 
for the mass range considered. We observe that the $K$-factor for $uu$ and $uc$ initial states decrease with $M_D$ 
while for $cc$ initial state it increases which is mainly because of the difference in the PDF distributions 
for the valence and sea quarks in the proton. Also note that the $K$-factors in the case of the vector sextet diquark are
larger compared to their corresponding values in the vector triplet case which is unlike that observed for the scalar 
diquarks. For the scalar diquarks there is a partial cancellation between the $C_F$ and $C_D$ terms, 
which gives a smaller $K$-factor for the sextet case compared to the antitriplet \cite{Han:2009ya},  while 
the $C_F$ and $C_D$ terms in the vector case come with the same sign. 
However other features such as a larger $K$-factor for the sea quarks compared to 
the valence quarks remains the same, as this comes from their PDF behaviour as the factorization scale varies.
\begin{table}[t]
\begin{center}
\begin{tabular}{|c|c|l|l|c|l|l|c|}
\hline
 & & \multicolumn{6}{|c|}{$\sqrt{s}=8~{\rm TeV}$}\\ 
\hline
 & & \multicolumn{3}{|c|}{$M_D=1~{\rm TeV}$} & \multicolumn{3}{c|}{$M_D=3~{\rm TeV}$}\\
\hline
qq  &State& LO & NLO &  $K_F$ & LO & NLO &  $K_F$\\
\hline
\hline
        & S      &$212^{+34.6}_{-27.3}$&$306^{+4.5}_{-12.1}$ &1.4 &$1.08^{+41.8}_{-30.3}$ &$1.47^{+10.4}_{-15.9}$ &1.3\\
        {$uu$}   
        & AT &$106^{+17.2}_{-13.9}$ &$144^{+3.9}_{-5.2} $ &1.3 &$0.54^{+25.2}_{-19.16}$ &$0.69^{+7.7}_{-8.9}$ &1.2\\
\hline
        & S      &$227^{+35.6}_{-27.8}$ &$334^{+5.2}_{-12.7} $ &1.4 & $0.67^{+43.1}_{-30.9}$ &$0.92^{+11.3}_{-16.5}$ &1.3\\
        {$ud$}
        & AT  &$113^{+18.0}_{-14.5}$ &$157^{+4.3}_{-5.6}$ &1.3 & $0.33^{+26.4}_{-19.84}$ &$0.43^{+8.3}_{-9.5}$ &1.2\\
\hline
        & S      &$57.3^{+36.6}_{-28.3} $&$86.0^{+5.8}_{-13.2} $ &1.4&$0.09^{+44.5}_{-31.4}$ &$0.13^{+12.3}_{-17.1}$ & 1.3\\
        {$dd$}    
        & AT &$28.6^{+18.8}_{-15.1} $ &$40.4^{+4.7}_{-6.0} $ &1.4 &$0.04^{+27.5}_{-20.5}$ &$0.06^{+8.9}_{-10.0}$ &1.2\\

\hline

        & S      & $0.89^{+36.6}_{-28.4} $&$1.40^{+8.3}_{-14.0} $ &1.5 &${5.46\times 10^{-5}}^{+45.1}_{-31.7}$ &${1.38\times 10^{-4}}^{+12.2}_{-16.8}$ &2.5\\
         {$ss$}
        & AT & $0.44^{+18.9}_{-15.2} $&$0.64^{+5.3}_{-6.1} $ &1.4&${2.73\times 10^{-5}}^{+28.1}_{-20.8}$ &${6.39\times 10^{-5}}^{+8.4}_{-9.4}$ &2.3\\
\hline

        & S      & $0.95^{+34.2}_{-27.3} $&$1.70^{+8.0}_{-13.4} $ &1.7 &${4.33\times 10^{-5}}^{+41.6}_{-30.2}$ &${1.92\times 10^{-4}}^{+11.1}_{-16.0}$ &4.4 \\
         {$sc$}
        & AT & $0.47^{+16.8}_{-14.0} $&$0.77^{+4.7}_{-5.3} $ &1.6 &${2.16\times 10^{-5}}^{+25.0}_{-19.1}$ &${8.87\times 10^{-5}}^{+7.5}_{-8.7}$ &4.1\\

\hline

        & S      & $0.24^{+31.7}_{-26.2} $&$0.51^{+7.6}_{-12.7} $ &2.1 &${8.65\times 10^{-6}}^{+38.2}_{-28.7}$ &${6.55\times 10^{-5}}^{+10.0}_{-15.1}$ &7.5\\
         {$cc$}
        & AT & $0.12^{+14.6}_{-12.6} $&$0.23^{+4.1}_{-4.6} $ & 1.9&${4.32\times 10^{-6}}^{+22.0}_{-17.3}$ &${3.01\times 10^{-5}}^{+6.6}_{-7.8}$ &6.9\\

\hline

        & S      & ${0.09}^{+24.1}_{-22.5} $&$0.19^{+6.0}_{-10.3} $ &2.0 &${3.31\times 10^{-6}}^{+31.3}_{-25.6}$ &${1.81\times 10^{-5}}^{+6.7}_{-12.4}$ &5.4\\
         {$bb$}
        & AT & ${0.04}^{+8.0}_{-8.3} $&${0.08}^{+2.1}_{-2.0} $ &1.7 &${1.65\times 10^{-6}}^{+15.9}_{-13.7}$ &${8.24\times 10^{-6}}^{+3.6}_{-5.0}$ &4.9\\
\hline

\end{tabular}
\caption{The LO and NLO cross sections (in pb) and K-factors for vector diquark production via different initial quark states at $\sqrt{s} = 8$ TeV. 
We give the cross sections  for both the sextet (S) and antitriplet (AT) diquarks. The uncertainties (in $\%$) given for the cross sections are due to the 
the choice of scale $Q=\mu$ and is obtained by varying the scale from $M_D/2$ to $2M_D$. We choose two reference values of 
the vector diquark mass $M_D= 1,3$ TeV and a fixed value for the coupling, $\lambda=1$.}         
\label{tab:8tev}                     
\end{center}
\end{table}

One of the primary reasons for calculating the higher-order corrections to a scattering process is to minimize the 
scale dependence on measurable observables such as cross sections, that would affect the event rate estimates 
at experiments.  We therefore make an estimate of the dependence of the choice of scale on the LO and NLO 
cross sections for the vector diquark production. To illustrate this we vary both the 
renormalization $\mu_R$ and factorization $\mu_F$ scale by a factor of two about the central scale $\mu =M_D$
keeping $\mu_R=\mu_F=\mu$ throughout. 
Note that the renormalization scale dependence of the leading order cross section is governed by the one-loop
running of the coupling parameter $\lambda$. Thus the scale dependence of the LO cross section has an 
uncertainty of ${\cal O}(\alpha_s)$. Although, while predicting the scale dependence of NLO cross section, 
we should use two-loop running of the coupling, leading to an uncertainty of ${\cal O}(\alpha_s^2)$: in absence of 
the two-loop result for running coupling we use Eq.~\ref{eq:running coupling} for predicting the renormalization 
scale dependence for both  the LO and NLO cross sections for the vector diquark production at LHC with $\sqrt{s}=13$ TeV. 
We plot our results in Fig. \ref{fig:13tev-uu-uc-cc-scale2}, where we can see clearly how the scale dependence of the 
NLO cross section is significantly reduced compared to the LO cross section. While the LO cross section varies between 
$\sim \pm 30\%$ for the vector sextet diquark for the three initial states $uu,~uc$ and $cc$ as $\mu$ varies between 
$M_D/2$ to $2 M_D$, the dependence is reduced to  $\sim \pm 10\%$ for the NLO cross sections. For the antitriplet 
vector diquark, the dependence is relatively less compared to the sextet, of about $\sim \pm (12-14)\%$ for the LO
cross sections which gets reduced to   $\sim \pm 4\%$ for the NLO result. Notice that the scale uncertainty in antitriplet 
case is much smaller than that in sextet case which is reduced further when the NLO results are included. 
This is because of the $C_D$ dependence (see Eq. \ref{eq:running coupling} and Eq. \ref{eqn:qq-NLO}) 
which is smaller for the antitriplet ($C_D=4/3$) compared to  the sextet ($C_D=10/3$). 

\begin{table}
 \begin{center}
 \begin{tabular}{|cc|c|c|c|c|c|c|}
\hline
  && \multicolumn{6}{|c|}{$\sqrt{s}=13~{\rm TeV}$}\\ 
   &         & \multicolumn{6}{|c|}{}\\
\hline
 & & \multicolumn{3}{|c|}{$M_D=1~{\rm TeV}$} & \multicolumn{3}{c|}{$M_D=3~{\rm TeV}$}\\
\hline
qq &State& LO & NLO &  $K_F$ & LO & NLO &  $K_F$\\
\hline
\hline
       & S      &$364^{+31.6}_{-25.8}$&$528^{+4.2}_{-11.3}$ &1.4&$6.91^{+35.5}_{-27.3}$ &$9.60^{+6.1}_{-12.5}$ &1.3\\
        {$uu$}   
       & AT &$182^{+14.6}_{-12.1}$ &$249^{+3.4}_{-4.4} $ &1.3&$3.45^{+19.7}_{-15.6}$ &$4.54^{+4.8}_{-6.2}$ &1.3\\
\hline 
       & S      &$443^{+32.3}_{-26.2}$&$660^{+4.7}_{-11.7}$ &1.4&$5.79^{+36.6}_{-27.8}$ &$8.12^{+6.8}_{-13.1}$ &1.4\\
        {$ud$}   
       & AT &$221^{+15.1}_{-12.6}$ &$310^{+3.7}_{-4.7} $ &1.3&$2.89^{+20.6}_{-16.2}$ &$3.83^{+5.2}_{-6.7}$ &1.3\\

\hline 
       & S      &$126^{+33.0}_{-26.5}$&$192^{+5.2}_{-12.1}$ & 1.5&$1.15^{+37.6}_{-28.3}$ &$1.62^{+7.5}_{-13.7}$ &1.4\\
        {$dd$}   
       & AT &$63.3^{+15.}_{-13.0}$ &$90.2^{+3.9}_{-4.9} $ &1.4&$0.57^{+21.5}_{-16.8}$ &$0.76^{+5.7}_{-7.1}$ &1.3\\

\hline  
       & S      &$4.75^{+32.1}_{-26.2}$&$7.50^{+7.6}_{-12.9}$ &1.5 &${3.73\times 10^{-3}}^{+38.8}_{-28.9}$ &${6.43\times 10^{-3}}^{+8.9}_{-14.1}$ &1.7\\
        {$ss$}   
      & AT &$2.37^{+14.9}_{-12.6}$ &$3.39^{+4.4}_{-4.9} $ &1.4&${1.86\times 10^{-3}}^{+22.6}_{-17.5}$ &${2.97\times 10^{-3}}^{+6.1}_{-7.1}$ &1.6\\
\hline  
      & S      &$5.82^{+30.1}_{-25.3}$&$9.91^{+7.7}_{-12.5}$ &1.7 &${3.31\times 10^{-3}}^{+36.4}_{-27.8}$ &${7.81\times 10^{-3}}^{+8.1}_{-13.3}$ &2.3\\
        {$sc$}   
      & AT &$2.91^{+13.2}_{-11.5}$ &$4.44^{+4.1}_{-4.4} $ &1.5&${1.65\times 10^{-3}}^{+20.4}_{-16.2}$ &${3.60\times 10^{-3}}^{+5.3}_{-6.3}$ &2.1\\

\hline  
      & S      &$1.72^{+27.9}_{-24.2}$&$3.23^{+7.6}_{-12.0}$ &1.8 &${7.26\times 10^{-4}}^{+33.9}_{-26.6}$ &${2.37\times 10^{-3}}^{+7.3}_{-12.5}$ &3.2\\
        {$cc$}   
      & AT &$0.86^{+11.3}_{-10.2}$ &$1.43^{+3.7}_{-3.8} $ &1.6&${3.63\times 10^{-4}}^{+18.2}_{-14.9}$ &${1.09\times 10^{-3}}^{+4.6}_{-5.6}$ &3.0\\

\hline  
      & S      &$0.73^{+20.8}_{-20.6}$&$1.34^{+6.8}_{-10.2}$ &1.8 &${2.80\times 10^{-4}}^{+27.9}_{-23.7}$ &${8.16\times 10^{-4}}^{+5.3}_{-10.3}$ &2.9\\
        {$bb$}   
      & AT &$0.36^{+5.1}_{-6.0}$ &$0.58^{+2.1}_{-1.7} $ &1.5&${1.40\times 10^{-4}}^{+12.9}_{-11.5}$ &${3.68\times 10^{-4}}^{+2.6}_{-3.3}$ &2.6\\
\hline  
 \end{tabular}
\caption{The LO and NLO cross sections (in pb) and K-factors for vector diquark production via different initial quark states at $\sqrt{s} = 13$ TeV.
All other choices are similar to that in Table \ref{tab:8tev}. }       
\label{tab:13tev}         
\end{center}
\end{table}
Note that we have till now chosen to illustrate our results with figures for only the 4/3 charged diquark production
that couple to the first two generations of the fermions. But we should also note here that the vector diquarks with 
the 2/3 and 1/3 charge can have substantial rates only affected by the initial PDF's of the contributing quarks. So 
to put the rate of production for the different vector diquarks in perspective we calculated all the modes that could contribute
to its production and present the LO and NLO cross sections in the relevant channels with scale uncertainties at $\sqrt{s}=8$ 
and $\sqrt{s}=13$ TeV run of LHC. To highlight the cross sections we have chosen two representative values of diquark mass 
$M_D= 1$ and $3$ TeV and fixed the coupling $\lambda=1$. We show the cross sections for LHC with $\sqrt{s}=8$ TeV and 
$13$ TeV in Table \ref{tab:8tev} and Table \ref{tab:13tev} respectively.  
We assume that the couplings of vector diquarks mediating quarks of different generations is suppressed. So out of the 15 possible 
combinations we only consider 7 combinations with no inter-generation vertices. One can clearly see that the valence 
quark contributions dominate,  with the $uu$ and $dd$ contributions being  a few orders of magnitude higher than $cc$ and $ss$ 
respectively for $M_D=1$ TeV in Table \ref{tab:8tev}. For the 3 TeV diquark, the difference in orders is nearly doubled. A similar 
behavior is seen in Table \ref{tab:13tev}. It is quite easy to understand that this happens due to the PDF's of the quarks in consideration
and the momentum fraction $x$ of the initial proton that they carry. However the notable thing to consider is the fact that due to  
\begin{figure}[b!]
\includegraphics[width=0.5\linewidth,height=0.45\linewidth]{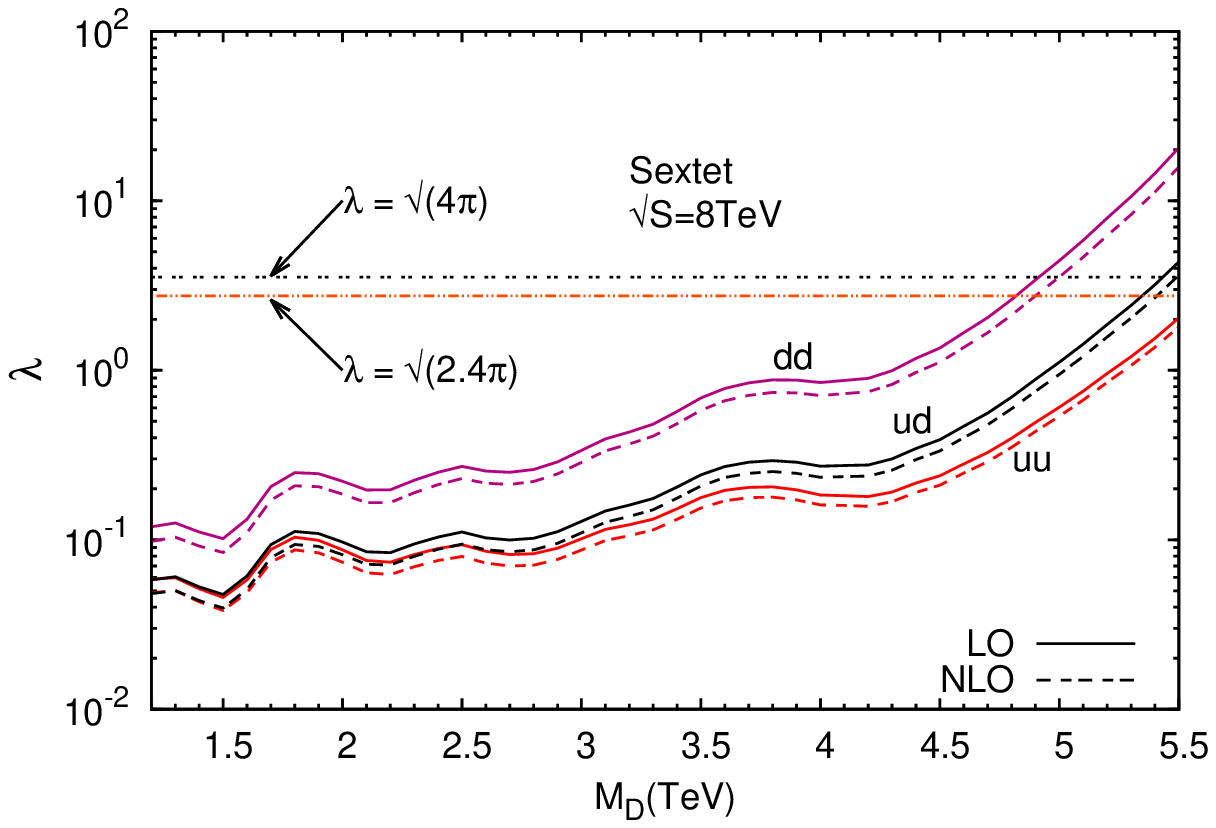}
\includegraphics[width=0.5\linewidth,height=0.45\linewidth]{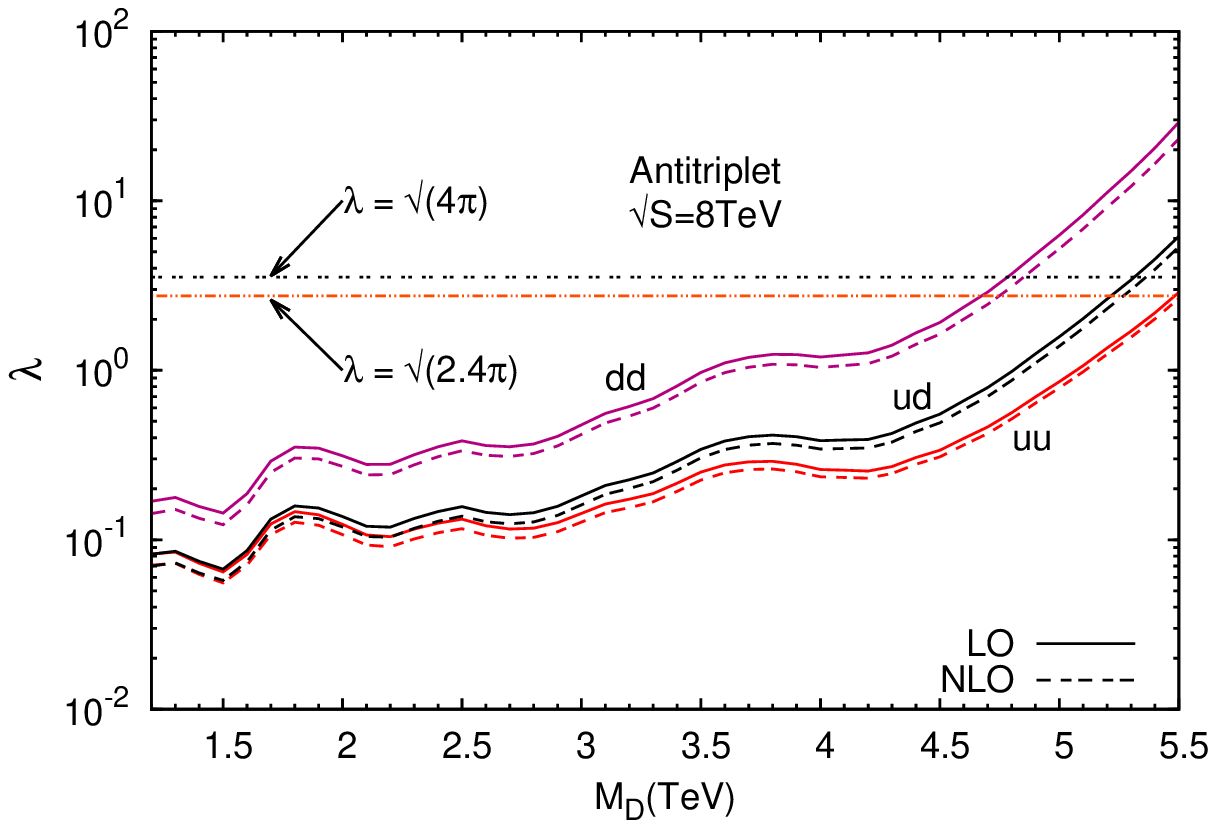}
\caption{The constraints on the mass $M_D$ and coupling $\lambda$ at $95\%$  C.L. for the sextet and
antitriplet vector diquark states at the LHC with $\sqrt{s} = 8$ TeV using the LO and NLO cross sections.
The values $\sqrt{4\pi}$ represents the perturbative limit for $\lambda$ while $\lambda=\sqrt{2.4\pi}$ 
gives the upper bound on the coupling for $\Gamma/M_D < 10\%$.}
\label{fig:cms-dijet-limit}
\end{figure}
quite small production cross sections for the diquarks produced through second generation quarks, even with order 1 coupling,
the mass limits on them would be considerably weaker compared to the diquarks coupling to the first generation. As we have already 
determined a rough order of magnitude by which the cross sections differ for the first and second generation vector diquarks, it 
would give us a comparative  idea of the limits on their coupling and mass from that derived for any one generation. We 
already have updated limits from dijet data by both ATLAS and CMS collaborations at the LHC \cite{Aad:2014bc, Khachatryan:2015jd}.
We use Ref. \cite{Khachatryan:2015jd} of the CMS collaboration to derive the limits on the vector diquark mass and coupling. 
The CMS collaboration has given the upper bound on the cross sections for different resonant mass values which can be 
compared with the parton-level resonant production cross section $(\sigma)$ times branching fraction $({\mathcal B})$ in the 
narrow-width approximation using $\sigma \mathcal{BA}$, where ${\mathcal A}$ is an acceptance factor $\sim 0.6$ \cite{Khachatryan:2015jd}.
We use this to derive limits for the vector diquark (both sextet and antitriplet) mass $M_D$ and its coupling $\lambda$ which interacts only
with the first generation quarks. As these would be contributions coming through the valence quarks with the largest rates, the limits on the diquark coupling to the 
second and third generation quarks would be much weaker. In Fig. \ref{fig:cms-dijet-limit} we show the $95\%$ C.L. constraints 
on the mass and couplings of the vector diquark produced through $uu, ud$ and $dd$ fusion using the dijet data from Ref. \cite{Khachatryan:2015jd}.
The plots illustrate that all values of $M_D$ and $\lambda$ which are above the curves are ruled out by the CMS dijet data at $95\%$ C.L..
Note that we assume that the vector diquark couples to only one pair of quarks. We also show the perturbative limit of $\lambda=\sqrt{4\pi}$ 
in the plots,  while $\lambda=\sqrt{2.4\pi}$  gives the upper bound on the coupling for $\Gamma/M_D < 10\%$.  As expected the strongest limits 
are for the $4/3$ charged diquark which couples to $uu$. The NLO corrections do modify the constraints to give slightly stronger limits 
compared to the LO results. For example, given a fixed value of the coupling $\lambda=0.5$ we find the $dd$ initiated 
LO result for the antitriplet vector diquark gives a lower bound of $M_D\simeq 3.03$ TeV whereas the NLO corrections improve the 
limit by about 100 GeV to $M_D \simeq 3.12$ TeV. The corresponding limits for the sextet vector diquark at LO ($M_D\simeq 3.32$ TeV) 
changes to  $M_D\simeq 3.42$ TeV at NLO. The corrections in the other modes are also found to be between $\sim$ 50-100 GeV. 
We have chosen not to show the effect of the associated scale 
uncertainties on the limits obtained. It should suffice to mention that the bounds using the LO cross sections would incorporate a 
much larger uncertainty band in the constraints compared to the NLO which is evident from the details given in Table \ref{tab:8tev} and 
Table \ref{tab:13tev}. Also note that as the cross section for the second generation induced productions are at least 2 or more orders 
of magnitude smaller for similar couplings, the limits on the couplings would be relaxed by a factor of $10$ or larger, allowing larger 
couplings for similar diquark mass. However one clearly finds a large parameter region still allowed for vector diquarks which should 
be explored at the upcoming run of LHC with $\sqrt{s}=13$ TeV.   
\section{Summary}\label{sec:concl}
In this work we have calculated the NLO QCD corrections to the vector diquark production at hadron colliders, namely the LHC. 
As colored particles are surely to be produced with large cross sections at hadron colliders, the discovery of any such state could be the 
first step towards discovering BSM physics at the LHC. Colored particles such as the vector diquark can mediate larger production rates
for dijet and multijet events. We show how the NLO corrections to the vector diquark production affects the cross sections for the 
sextet and antitriplet representations. As the vector diquark couplings to the quark pair can be generation dependent, we find that 
valence quark processes have $K$-factors in the range of 1.5 to 1.3 for a mass range of 0.5-1.5 TeV which decrease as we 
go higher in mass. The $sea$ quark initiated production modes are found to have increasing values of the 
$K$-factor as the diquark mass is increased. We also find that unlike the scalar diquarks,
the sextet vector diquark has larger NLO corrections compared to the antitriplet. 
We also illustrate the scale uncertainties in the 
cross section for both the sextet and antitriplet vector diquarks and find that the sextet vector diquark exhibits bigger scale 
uncertainty at LO compared to the antitriplet. The NLO corrected 
cross sections for both cases are found to show much lesser dependence on the scale variation. We also calculate the NLO corrections 
to the width of the vector diquark decaying to a pair of quarks. As a narrow-width approximation is considered large corrections to the 
width can affect predictions for relevant final states. We find that the $K$-factor for decay width of the sextet diquark is around $1.08-1.1$
while it is around $1.05$ for the antitriplet which is relatively smaller than that for the production cross section. However the scale uncertainties
are relatively large for the decay width which get reduced by taking the NLO corrected widths. 

We have calculated cross sections for the vector diquark production at LHC with $\sqrt{s}=8$ and $13$ TeV arising from different generation quarks.
We use the dijet data from the CMS collaboration for LHC with $\sqrt{s}=8$ TeV  to put limits on the vector diquark mass and its coupling. We find that  
a large parameter region is still allowed for vector diquarks which should be explored at the upcoming run of LHC at $\sqrt{s}=13$ TeV. The current 
limits by the LHC experiments on the resonant particles include scalar diquarks but do not include vector diquarks. We have shown that using the 
same data one could also search for the vector diquarks and give search limits for such particles.   
\acknowledgments{ We would like to thank  M.C. Kumar, M.K. Mandal, V. Ravindran and E. Vryonidou for fruitful discussions. We thank 
IACS, Kolkata for hospitality during the LHCDM-2015 Workshop and AS would also like to thank CP3-Louvain, Belgium for hospitality while part of 
the work was carried out. 
The work of KD, SKR and AS is partially supported by funding available from the Department of Atomic Energy, 
Government of India, for the Regional Centre for Accelerator-based Particle Physics (RECAPP), Harish-Chandra Research Institute. 
The work of SM is partially supported by CSIR SRA under Pool Scheme (No.13(8545-A)/Pool-2012). }
 
 \newpage
 \appendix

 \section{Feynman rules}\label{app:FR}
 
 The interaction Lagrangians given in Eqns~\ref{eq:Lagrn-QQD} and \ref{eq:Lagrn-GDD} give the following Feynman rules 
 (all momenta incoming) :
 
 \begin{itemize}
  
   \item    
  ${ \bf \bar{q^c_a}(p_1)q_b '(p_2)V_i^\mu(p_3) :} $ ~~~~~~~~~~~ $ \frac{i \lambda_{qq'}}{\sqrt {1+\delta_{qq'}}} \gamma_\mu (K^i_{ab}P_\tau-\delta_{qq'}K^i_{ba}P_{\tau '} )$\\
  where $P_\tau(P_{\tau'})$ can be $P_{L/R} (P_{R/L})$.
  
 \item
  ${ \bf V_i^{\mu_1}(p_1)V_j^{*\mu_2}(p_2)G^{A,\mu_3}(p_3) :}$ ~~ $ -i g_sT^A_{ji} [ g^{\mu_1\mu_2}(p_1-p_2)^{\mu_3}  + g^{\mu_2\mu_3}(p_2-p_3)^{\mu_1} \\
 \hspace*{3.8in}+ g^{\mu_3\mu_1}(p_3-p_1)^{\mu_2}] $

 \end{itemize}

\section{One-loop scalars}\label{app:OLS}

Here we list various tadpole ($A_0$), bubble ($B_0$) and triangle ($C_0$) scalar
integrals required in the calculation of virtual corrections in sections \ref{sec:pheno} 
and \ref{sec:decay}. 
For simplicity we take out the universal one-loop factor from these integrals which arise
in DR and use the following notation,
\begin{equation}
 I_0 = \frac{i}{16\pi^2} \frac{(4\pi \mu^2)^\epsilon}{\Gamma(1-\epsilon)}~ \tilde I_0.
\end{equation}

We have labeled the UV and IR singularities of scalar integrals explicitly in our calculations. In DR, 
$\epsilon_{\rm UV} = \epsilon_{\rm IR} = \epsilon$.

\begin{eqnarray}
&\tilde A_0(m^2)&= (m^2)^{(1-\epsilon)}\Big[ \frac{1}{\epsilon_{\rm UV}}+1 \Big] \\
 &\tilde B_0(s;0,0) &= \frac{1}{(-s)^\epsilon} \Big[ \frac{1}{\epsilon_{\rm UV}}+2 \Big] \\
 &\tilde B_0(0;0,m^2) &= \frac{1}{(m^2)^\epsilon} \Big[ \frac{1}{\epsilon_{\rm UV}}+1 \Big] \\
&\tilde B_0(0;0,0) &= \frac{1}{(\mu^2)^\epsilon}\Big[ \frac{1}{\epsilon_{\rm UV}}- \frac{1}{\epsilon_{\rm IR}} \Big] \\
&\tilde B_0(m^2;0,m^2) &= \frac{1}{(m^2)^\epsilon} \Big[ \frac{1}{\epsilon_{\rm UV}}+2 \Big] \\
 &\frac{\partial}{\partial s} \tilde B_0(s;0,m^2)|_{s=m^2} &= 
(m^2)^{(-1-\epsilon)}\Big[ -\frac{1}{ 2 \epsilon_{\rm IR}} -1 \Big]\label{eq:derivative-B0} \\
&\tilde C_0(0,0,s;0,0,0) &=  \frac{1}{(-s)^\epsilon} 
\Big[ \frac{1}{s} \Big(\frac{1}{\epsilon_{\rm IR}^2}\Big) \Big] \\
&\tilde C_0(0,0,m^2; 0,0,m^2) &= (m^2)^{(-1-\epsilon)}
\Big[ -\frac{1}{ 2 \epsilon_{\rm IR}^2} -\frac{\pi^2}{12} \Big]
\end{eqnarray}
The derivative of bubble function in Eq.~\ref{eq:derivative-B0} is 
used in the calculation of $Z_2^q$ and $Z_2^D$.
 
 \section{Plus function}\label{app:PF}
 For a function $f(x)$, singular at $x=1$, and a smooth 
 function $g(x)$, the {\it plus function} is defined by the 
 following relation,
 \begin{equation}
  \int_{0}^{1} dx f_{+}(x) g(x) = \int_{0}^{1} f(x)[g(x)-g(1)].
 \end{equation}
 Few {\it plus function} related identities which have been very useful in the 
 calculation of real corrections are,
 \begin{eqnarray}
   \int_{a}^{1} dx f_{+}(x) g(x)&=& \int_{a}^{1} dx f(x)[g(x)-g(1)]-g(1)\int_{0}^{a} dx f(x) \\
\frac{1}{(1-\tau)^{(1+2\epsilon)} } &=&  \frac{1}{(1-\tau)_+} 
                                    -2\epsilon \Big[ \frac{\ln(1-\tau)}{1-\tau}\Big]_+ 
                                    -\frac{1}{2\epsilon}\delta(1-\tau) \\
  \frac{f(x)}{(1-\tau)_+} &=& \Big[ \frac{f(x)}{(1-\tau)}\Big]_+ 
                           + \delta(1-\tau) \int_0^1 dz~ \frac{f(z)-f(1)}{1-z} 
 \end{eqnarray}

 \section{ ${\cal O}(\alpha_s)$ Correction to scalar diquark decay width}\label{app:SDW-NLO}
  The NLO QCD correction to the decay width for scalar diquark decaying into a pair of light jets is given by,
 \begin{eqnarray}
  \Gamma_{\rm NLO} = \Gamma_0 \Big\{ 1 + \frac{\alpha_s}{2\pi} \Big[ C_D\Big(\frac{5}{2} -\frac{2}{3}\pi^2\Big) 
   + C_F \Big( 3~{\rm ln}\Big(\frac{\mu_R^2}{M_D^2}\Big) + \frac{17}{2} \Big)\Big]  \Big\}, 
 \end{eqnarray}
 where, the LO decay width $\Gamma_0$, is given by 
  \begin{equation}
   \Gamma_{0} = \frac{\lambda^2}{16\pi}~ M_D.
  \end{equation}
Note that the $C_F$ part is exactly the same as one gets in the NLO QCD calculation  
 of $H \to b {\bar b}$ decay width~\cite{Braaten:1980xx}. We have used the following 
 interaction Lagrangian for the scalar diquark ($\Phi_i$) case,
 \begin{equation}
  {\cal L}^\phi = \frac{\lambda}{(1+\delta_{qq'})} \; [  \Phi_i \bar{q^c_a} K^i_{ab} P_\tau q'_b + {\rm h.c.} ] + (D_\mu\Phi_i)^\dagger (D^\mu\Phi_i) - M_D^2 \Phi_i^\dagger \Phi_i.
 \end{equation}
It should be noted that the coupling of the scalar diquark with two same flavor quarks is zero in antitriplet case.

 \section{Useful relations}
 Some of the relations among color factors that we have used to simplify various expressions in 
 sections~\ref{sec:pheno} and \ref{sec:decay}, are given below. For a more 
 complete list one may refer to Ref.~\cite{Han:2009ya}.
 \begin{eqnarray}
 t^A_{ab}t_{A,bc} &=& C_F \delta_{ac} \\
 T^A_{ij}T_{A,jk} &=& C_D \delta_{ik} \\
{\rm Tr}(K^i{\bar K}^i) &=& N_D \\
{\rm Tr}(K^i t^A t_A {\bar K}^i) &=& C_F N_D \\
{\rm Tr}(K^i t^A {\bar K}^i(t_A)^T) &=& \pm \frac{1}{2} C_F N_C \\
T^A_{ij}{\rm Tr}(K^j t^A {\bar K}^i)  &=&  T^A_{ij}{\rm Tr}({\bar K}^i (t^A)^T K^j  ) 
=  \frac{1}{2} C_D N_D \\
  \pm C_F N_C &=& -2C_F N_D + C_D N_D.
\end{eqnarray}
In the above $t^A_{ab}$ are the $SU(3)_C$ generators in fundamental representation while 
$T^A_{ij}$ are the generators in the diquark representation of $SU(3)_C$.

To calculate the real corrections to the 2-body decay of the diquark, the following relation has been used in simplifying 
the three body phase space integration in $n=4-2\epsilon$ dimensions.
\begin{eqnarray}
\Gamma(2z) = \frac{2^{2z-1}}{\sqrt{\pi}}\Gamma(z)\Gamma\Big(z+\frac{1}{2}\Big).
 \end{eqnarray}

\end{document}